\documentclass[twocolumn]{aastex62}

\usepackage{multirow}

\graphicspath{{./}{figures/}}

\received{June 21, 2018}
\revised{August 27, 2018}
\accepted{\today}
\submitjournal{ApJ}

\shorttitle{Magnetic shielding of soft protons}
\shortauthors{Fioretti et al.}

\begin{document}

\title{Magnetic shielding of soft protons in future X-ray telescopes: the case of the ATHENA Wide Field Imager.}

\correspondingauthor{V. Fioretti}
\email{valentina.fioretti@inaf.it}

\author[0000-0002-6082-5384]{Valentina Fioretti}
\affil{INAF OAS Bologna, Via P. Gobetti 101, 40129, Bologna, Italy}

\author{Andrea Bulgarelli}
\affil{INAF OAS Bologna, Via P. Gobetti 101, 40129, Bologna, Italy}

\author{Silvano Molendi}
\affil{INAF IASF Milano, Via E. Bassini 15, 20133 Milano, Italy}

\author{Simone Lotti}
\affil{INAF IAPS, Via del Fosso del Cavaliere, 100, 00133 Roma, Italy}

\author{Claudio Macculi}
\affil{INAF IAPS, Via del Fosso del Cavaliere, 100, 00133 Roma, Italy}

\author{Marco Barbera}
\affil{Palermo University, Dep. of Physics and Chemistry, Via Archirafi 36, 90123 Palermo, Italy}

\author{Teresa Mineo}
\affil{INAF IASF Palermo, Via Ugo La Malfa 153, 90146 Palermo, Italy}

\author{Luigi Piro}
\affil{INAF IAPS, Via del Fosso del Cavaliere, 100, 00133 Roma, Italy}

\author{Massimo Cappi}
\affil{INAF OAS Bologna, Via P. Gobetti 101, 40129, Bologna, Italy}

\author{Mauro Dadina}
\affil{INAF OAS Bologna, Via P. Gobetti 101, 40129, Bologna, Italy}

\author{Norbert Meidinger}
\affil{Max-Planck-Institut f\"{u}r extraterrestrische Physik, Giessenbachstrasse 1, 85748 Garching, Germany}

\author{Andreas von Kienlin}
\affil{Max-Planck-Institut f\"{u}r extraterrestrische Physik, Giessenbachstrasse 1, 85748 Garching, Germany}

\author{Arne Rau}
\affil{Max-Planck-Institut f\"{u}r extraterrestrische Physik, Giessenbachstrasse 1, 85748 Garching, Germany}



\begin{abstract}

Both the interplanetary space and the Earth magnetosphere are populated by low energy ($\leq300$ keV) protons that are potentially able to scatter on the reflecting surface of Wolter-I optics of X-ray focusing telescopes and reach the focal plane. This phenomenon, depending on the X-ray instrumentation, can dramatically increase the background level, reducing the sensitivity or, in the most extreme cases, compromising the observation itself. The use of a magnetic diverter, deflecting protons away from the field of view, requires a detailed characterization of their angular and energy distribution when exiting the mirror. We present the first end-to-end Geant4 simulation of proton scattering by X-ray optics and the consequent interaction with the diverter field and the X-ray detector assembly, selecting the ATHENA Wide Field Imager as a case study for the evaluation of the residual soft proton induced background. We obtain that, in absence of a magnetic diverter, protons are indeed funneled towards the focal plane, with a focused Non X-ray Background well above the level required by ATHENA science objectives ($5\times10^{-4}$ counts cm$^{-2}$ s$^{-1}$ keV$^{-1}$), for all the plasma regimes encountered in both L1 and L2 orbits. These results set the proton diverter as a mandatory shielding system on board the ATHENA mission and all high throughput X-ray telescopes operating in the interplanetary space. For a magnetic field computed to deflect 99\% of the protons that would otherwise reach the WFI, Geant4 simulations show that this configuration, in the assumption of a uniform field, would efficiently shield the focal plane, yielding a residual background level of the order or below the requirement.

\end{abstract}



\section{Introduction}
Low energy protons ($\leq300$ keV, so-called soft protons), populating the interplanetary space and the Earth magnetosphere, can enter the field of view of X-ray focusing telescopes and then be funneled towards the focal plane by scattering at grazing angles with the mirror surface. This phenomenon was discovered after the damaging of the Chandra/ACIS front-illuminated CCDs in 1999 during its first passages through the radiation belt \citep{2000SPIE.4140...99O}. The damage was soon minimized by switching off the CCDs and moving them from the focal position \citep{2007SPIE.6686E..03O}.
Blocking filters protect XMM-Newton focal plane below an altitude of 40000 km, but above this limit soft protons induce sudden flares in the background count rate of the EPIC instruments. These events last from hundreds of seconds to hours and can hardly be disentangled from X-ray photons, causing the loss of large amounts (30-40\%) of observing time \citep{2017ExA....44..297M}. While telescopes operating in low Earth orbit are shielded by the Earth geomagnetic cut-off, the performance of future X-ray focusing telescopes orbiting in the interplanetary space can potentially suffer from soft proton induced background events. Examples of such missions are the ESA next large class ATHENA \citep{2013arXiv1306.2307N}, to be launched in 2030, or the eROSITA X-ray telescope on board the Russian/German Spectrum Roentgen Gamma Mission \citep{2014SPIE.9144E..1TP}, to be launched in 2019.
\\
The large effective area (1.4 m$^{2}$ at 1 keV) makes the minimization of soft proton contamination a key challenge for the fulfillment of ATHENA's science objectives. A possible shielding solution is placing an array of magnets (a magnetic diverter) between the optics and the focal plane, able to deflect charged particles away from the instruments field of view. 
X-ray telescopes on board Chandra, XMM-Newton, and Swift are already equipped with diverters deflecting the electrons populating the radiation belts (see e.g \citet{wil00}).
A proton diverter however, because of the $\sim2000$ times higher mass of the particle with respect to electrons, imposes a dedicated trade-off among the required magnetic field, the mass budget, and the impact on surrounding instruments.
\\
The ATHENA Wide Field Imager (WFI) aims, among the many scientific objectives \citep{2016SPIE.9905E..2BR}, to perform X-ray surveys of the high-z sky, populated by faint point sources, and to map the diffuse and faint thermal emission in clusters of galaxies. A low instrumental background is mandatory for the achievement of those science objectives \citep{wfi_bkg2018}. Because of this requirement and its large field of view ($40'\times40'$), the WFI is the best case study for the evaluation of the soft proton induced X-ray background and the shielding efficiency of a magnetic diverter placed in front of it. For the first time we present an end-to-end simulation of the soft proton induced background, including (i) the collection of all plasma regimes encountered in L2 (Sec. \ref{sec:l2}), (ii) the interaction of protons with the mirror (Sec. \ref{sec:sca}), (iii) the consequent interaction with optical blocking filter in the field of view, the surrounding structure, and the detector itself (Sec. \ref{sec:wfi}) and (iv) the impact of a magnetic diverter in deflecting proton tracks from the WFI (Sec. \ref{sec:div}). The evaluation of the ATHENA diverter efficiency is one of the products of the ESA AREMBES (ATHENA Radiation Environment Models and X-ray Background Effects Simulators) project\footnote{http://space-env.esa.int/index.php/news-reader/items/AREMBES.html}, that will deliver to the science community a full modeling of the ATHENA space radiation environment and a Geant4-based framework, including the full ATHENA mass model, for the simulation of the Non X-ray Background (NXB).


\section{Low energy protons in L2}\label{sec:l2}
The baseline for ATHENA is a periodic halo orbit around L2, the second Lagrange point of the Sun-Earth system, located 1.5 million kilometres from the Earth in the opposite direction to the Sun and where no X-ray telescope has ever flown.
A halo orbit is a particular type of Lissajous trajectories around L2 where the in-plane and out-of-plane frequencies are the same.
An L2 orbit, because of the constant Sun direction and the lack of passages in the Earth shadow, ensures a stable thermal environment to the instruments while achieving an almost constant view of the sky. 
\\
The interaction of the solar wind with the Earth magnetic field compresses the magnetosphere at the Sun side while generates a long tail of trapped particles, the magnetotail, at the opposite side. The L2 point is placed inside the Earth's magnetotail, and a spacecraft orbiting around it will encounter different regions of the tail, each characterized by different temperature and magnetic conditions. Depending on the halo orbit width, the spacecraft could also be directly exposed to the solar wind. 
The AREMBES project has achieved a detailed characterization of the L2 environment \citep{WP2_report1, WP2_report2} with the aim of modeling both the low and high energy particles encountered by the spacecraft. We focus here on the soft proton fluxes encountered along the orbit while crossing the magnetotail trapped population or the interplanetary outer regions exposed to the solar wind.
Since the ATHENA scientific requirement \citep{ATHENA_req} on the soft proton induced background must not be exceeded on 90\% of the mission time, all fluxes refer to a cumulative fraction of 90\%, i.e. the maximum flux to be encountered for 90\% of the time.

 \begin{figure*}
 \includegraphics[trim={4cm 3cm 1cm 1cm},clip,width=0.53\textwidth]{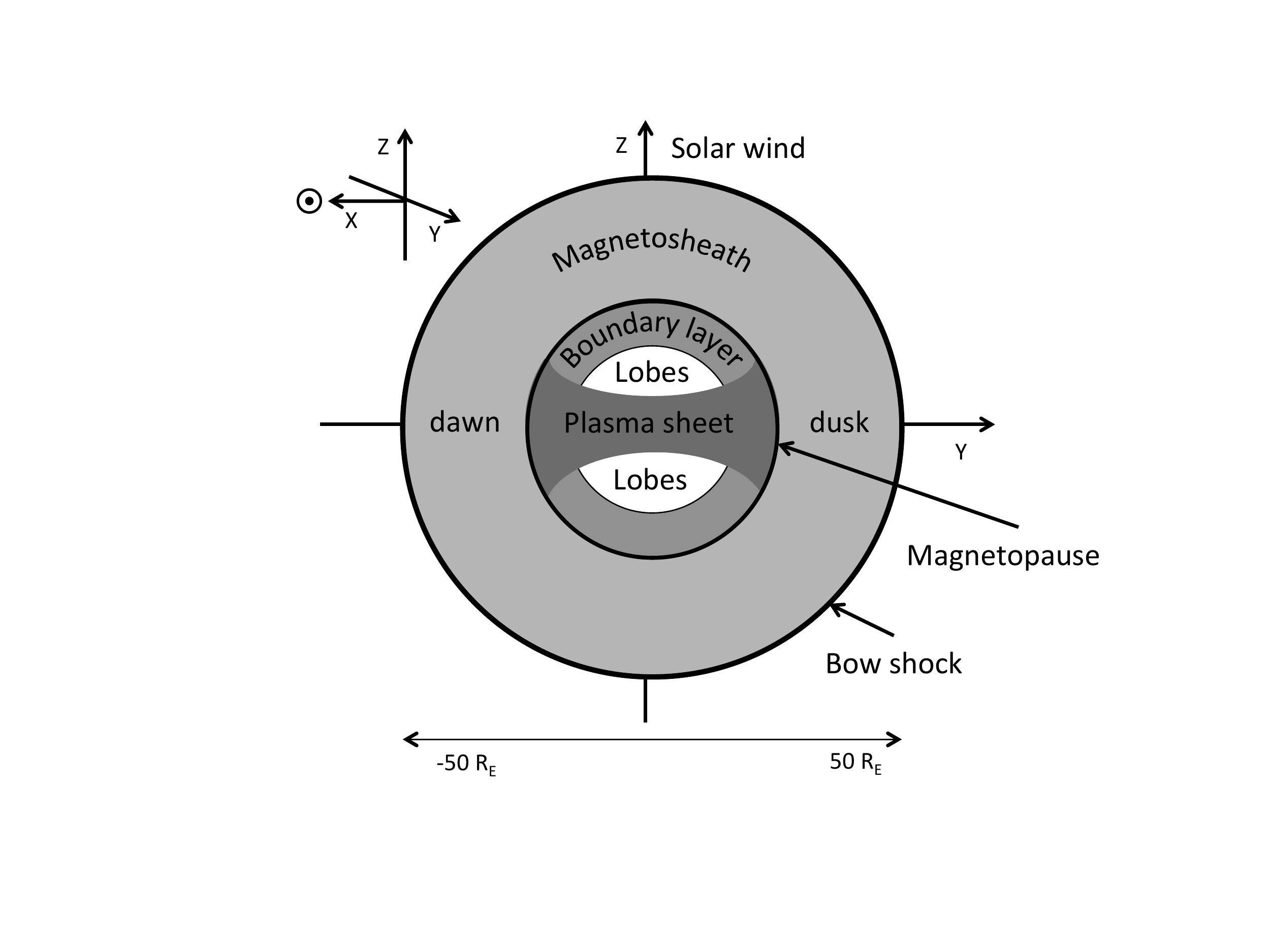}
  \includegraphics[trim={1cm 7cm 2cm 8cm},clip,width=0.47\textwidth]{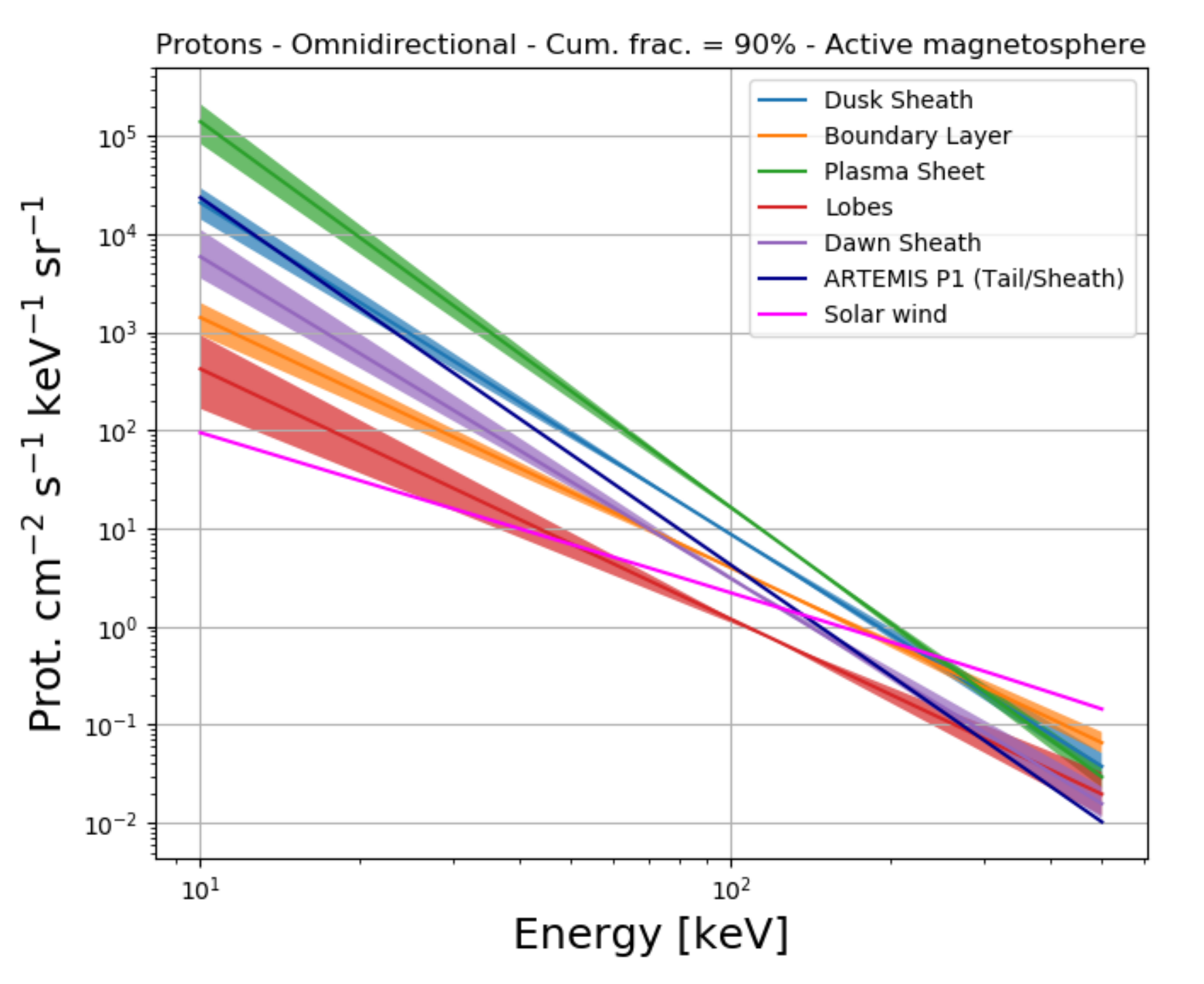}
   \caption 
   { \label{fig:input}\textit{Left panel}: plasma regimes composing the tail of the Earth magnetosphere in the Y-Z plane of the GSM system. \textit{Right panel}: the low energy proton spectral distribution for each plasma regime, averaged over an omni-directional flux, corresponding to the maximum flux expected to encounter for 90\% of the mission time in an active magnetosphere.}
   \end{figure*} 
   
\subsection{Interplanetary solar wind}
The solar wind soft proton contribution is based on the long term monitoring of almost two solar cycles performed by the EPAM/LEMS instrument (0.047 -- 4.8 MeV, \citet{gol98}) on board the NASA ACE satellite in the L1 region. 
Since the same fluxes and time variability are measured by the WIND spacecraft \citep{WP2_report2} through the L2 region, we can consider the L1 proton measurements a good estimate of the solar wind powered proton fluxes expected for the ATHENA spacecraft in L2. We use as reference flux the spectral fit in the ($\sim50 - 200$) keV energy range reported in \citet{lotti_2018_submitted} for a maximum cumulative fraction of 90\%. The estimates for solar wind powered, soft proton induced X-ray background in L2 would also apply for the case of ATHENA orbiting in L1.

\subsection{Magnetotail plasma regimes}
The distant tail of the Earth magnetosphere has been extensively mapped by the JAXA/NASA Geotail mission \citep{1994GeoRL..21.2871N}, flown from October 1992 to November 1994. The AREMBES modeling of the magnetotail plasma regimes is based on the Geotail data catalogue \citep{1998JGR...10323503E, 1998JGR...10323521C} after excluding all the time intervals corresponding to solar events. Magnetotail structure and inner morphology are highly dynamic because of solar and geomagnetic activity:  plasma regimes are identified with selection criteria based on magnetic field parameters, particle spectra, and angular distributions. According to \citet{1998JGR...10323503E}, the variability is such that a spacecraft placed in any point of the tail would cross almost all regimes. The cross section of the magnetotail in the Y-Z plane, listing all regimes under study, is shown in Fig. \ref{fig:input} (left panel). The coordinate system in the picture follows the GSM (Geocentric Solar Magnetic) system \citep{2017SSRv..206...27L}, with the X-axis from the Earth center to the Sun and the Z-axis along the northern magnetic pole. The magnetosheath (or \textit{Sheath}) lies behind the bow shock boundary layer, where the solar wind encounters the magnetosphere. The magnetosheath plasma, hotter and more turbulent than the solar wind one \citep{2013JGRA..118.4963D}, extends to the magnetopause layer. Since the interplanetary magnetic field lines are frozen in the solar wind plasma, their inclination is quasi-parallel (dawn side) or quasi-perpendicular (dusk side) to the normal from the shock surface, thus causing a difference in the magnetosheath composition according to the dawn/dusk side. 
Within the magnetopause, the tail structure includes the plasma sheet, a sheet-like region in the equatorial magnetotail characterized by low magnetic fields and enriched by H ions of solar origin \citep{1967JGR....72..113B}, encircled by the cold plasma of the lobes. The boundary layers are placed between the lobes and the magnetopause and are mainly composed by low energy, low field plasma. The magnetosheath regime is most probable up to $\pm50$ R$_{\rm E}$ (Earth radii) in the Y$_{\rm GSM}$ axis according to Geotail observations. Since its orbit reaches a distance from the Earth of $\sim200$ R$_{\rm E}$, while the L2 orbit center is placed at $235$ R$_{\rm E}$, the magnetosheath width could be larger at the ATHENA orbital region.
 \begin{figure*} 
   \includegraphics[trim={4cm 0cm 4cm 0cm},clip,width=0.47\textwidth]{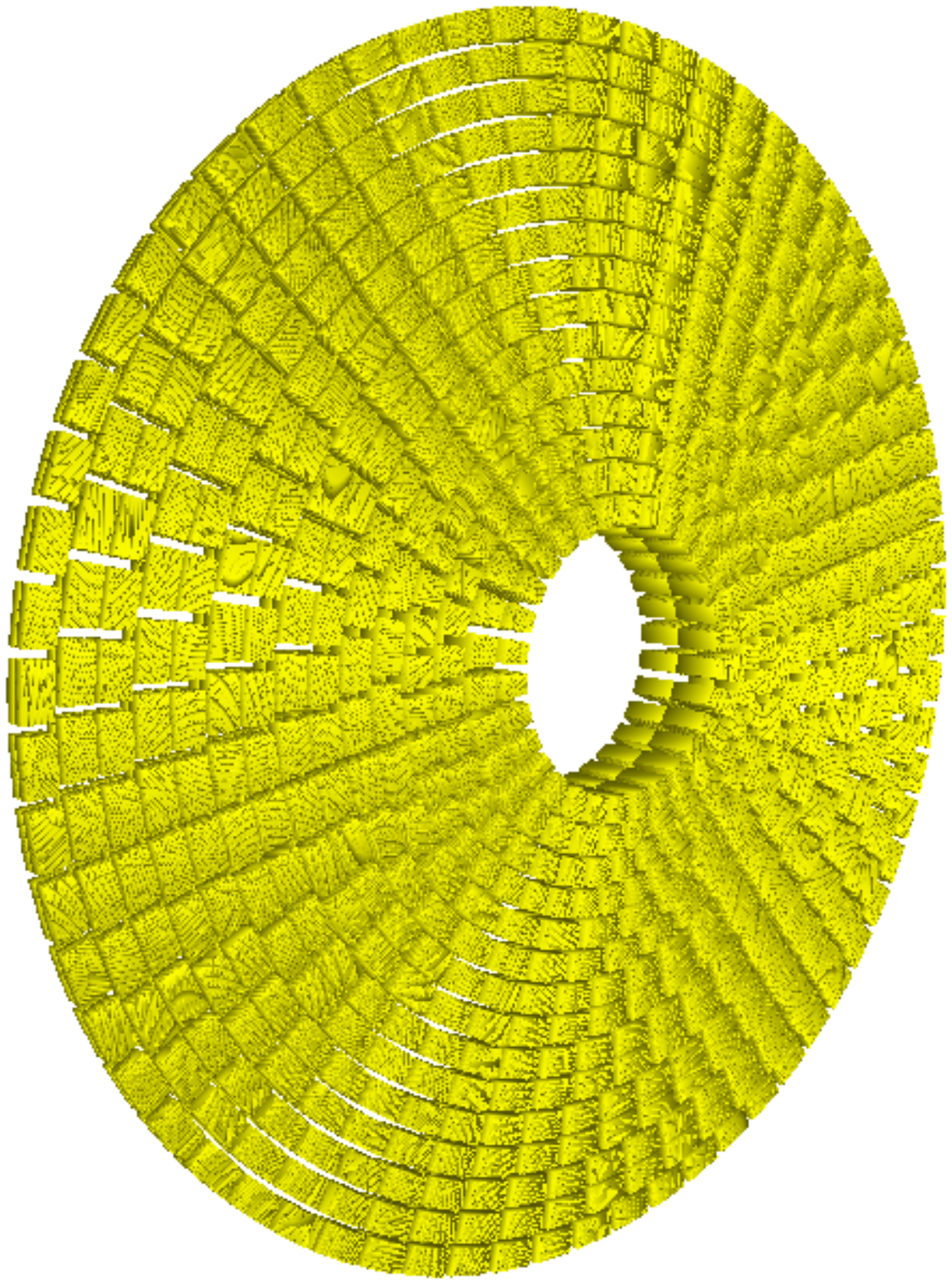}
   \includegraphics[width=0.53\textwidth]{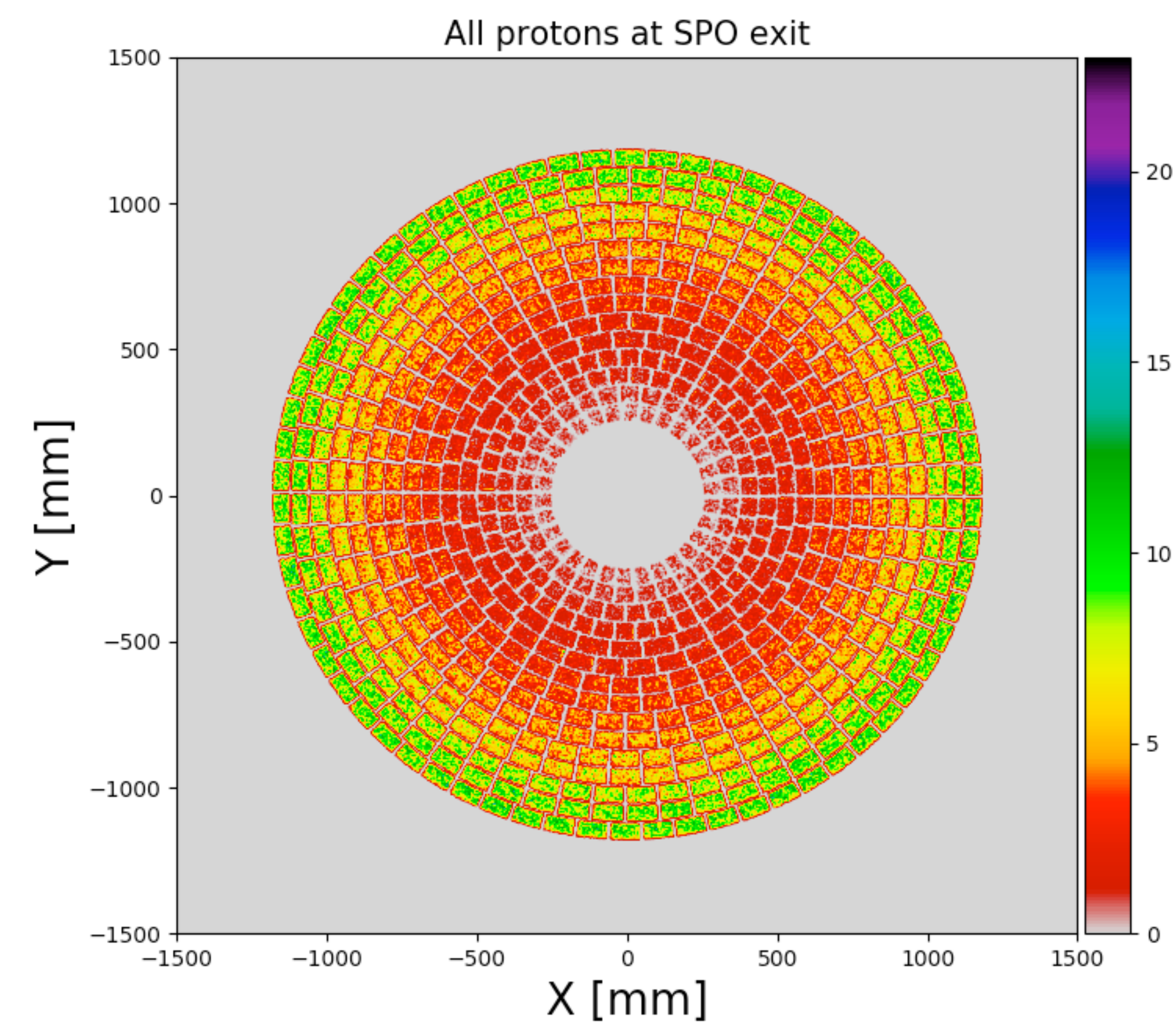}
   \caption 
   { \label{fig:spo}Geant4 mass model (left panel) of ATHENA SPO for the current design and the spatial distribution ($5\times5$ mm$^2$ binning) of protons scattered on the Si pores and exiting the mirror (right panel). The color bar unit is protons.}
   \end{figure*} 
\subsection{Soft proton flux in L2}
The soft proton spectral distribution for each listed plasma regime uses the power-law best fit from the AREMBES study \citep{WP2_report2}, with a confidence domain $<2$ times the root mean square (r.m.s.) of the fit. Although the study provides fluxes selected for different directions, given the high variable nature of the environment, we take as reference the averaged omni-directional flux. Geotail data for protons start at 58 keV but we extend all models down to 10 keV, assuming that the same distribution at low energies. We perform the extrapolation because we expect the 10 -- 100 keV proton population to be the major contributor in the WFI background (see Sec. \ref{sec:res} for details). As for the solar wind model, all fluxes refer to a cumulative fraction of 90\%.
The geomagnetic activity is the primary contributor to the ion intensity fluctuation in the magnetotail \citep{1994GeoRL..21.3015J}, hence we refer to the higher plasma regimes reported for an active magnetosphere (Auroral Electrojet Index AE $>125$ nT).
\\
Fig. \ref{fig:input} (right panel) reports, in addition to the listed regimes from the Geotail analysis and the ACE data, the soft proton power-law spectrum obtained from NASA Artemis P1 observations in the 30 -- 80 keV energy range in the magnetotail/magnetosheath region \citep{2017ITPS...45.1965B}, for a cumulative fraction of 90\% and with the model extended from 10 to 500 keV.
\\
The halo orbit for ATHENA is yet to be defined, but we report as reference the Herschel spacecraft large halo orbit in L2 \citep{Hmanual}, with a maximum radius of $\sim50$ R$_{\rm E}$ in the Z$_{\rm GSM}$-axis and $\sim100$ R$_{E}$ in the Y$_{\rm GSM}$-axis. This orbit, as also fully reported by \citet{lotti_2018_submitted}, would mostly expose the ATHENA telescope to the plasma composing the magnetosheath and the solar wind. Following a conservative approach, the simulated soft proton flux at the ATHENA Silicon Pore Optics (SPO) entrance is the plasma sheet population, the plasma regime with the highest intensity in the magnetotail, but the soft proton induced background on the WFI will be also computed for all the other regimes after proper scaling.

\section{Soft proton scattering}\label{sec:sca}
Soft protons scatter on the mirror surface, and the resulting energy, angular distribution, and scattering efficiency affect the residual proton flux reaching the focal plane. The end-to-end Geant4 simulation of the soft proton induced background on the WFI, comprising both the proton interaction with the optics and the WFI assembly, requires the accurate modeling of the ATHENA SPO structure and their physics of interaction. 
\\
The Geant4 \citep{g4_1, g4_2, 2016NIMPA.835..186A} toolkit, initially developed by CERN and then maintained by a large collaboration, is a C++ based particle transport code for the simulation of high energy experiments at particle accelerators and then extended to other scientific communities and lower energies (sub-keV scale). All simulations presented here, based on the Geant4 10.3 release, are obtained with the BoGEMMS (Bologna Geant4 Multi-Mission simulator) simulation framework \citep{2012SPIE.8453E..35B}, a customizable and astronomy-oriented Geant4-based simulation tool for the evaluation of the scientific performance (e.g. background spectra, effective area) of high energy experiments, with particular focus on X-ray and gamma-ray space telescopes. 
 \begin{figure*} 
 \centering
   \includegraphics[trim={1cm 7cm 1cm 8cm},clip,width=0.7\textwidth]{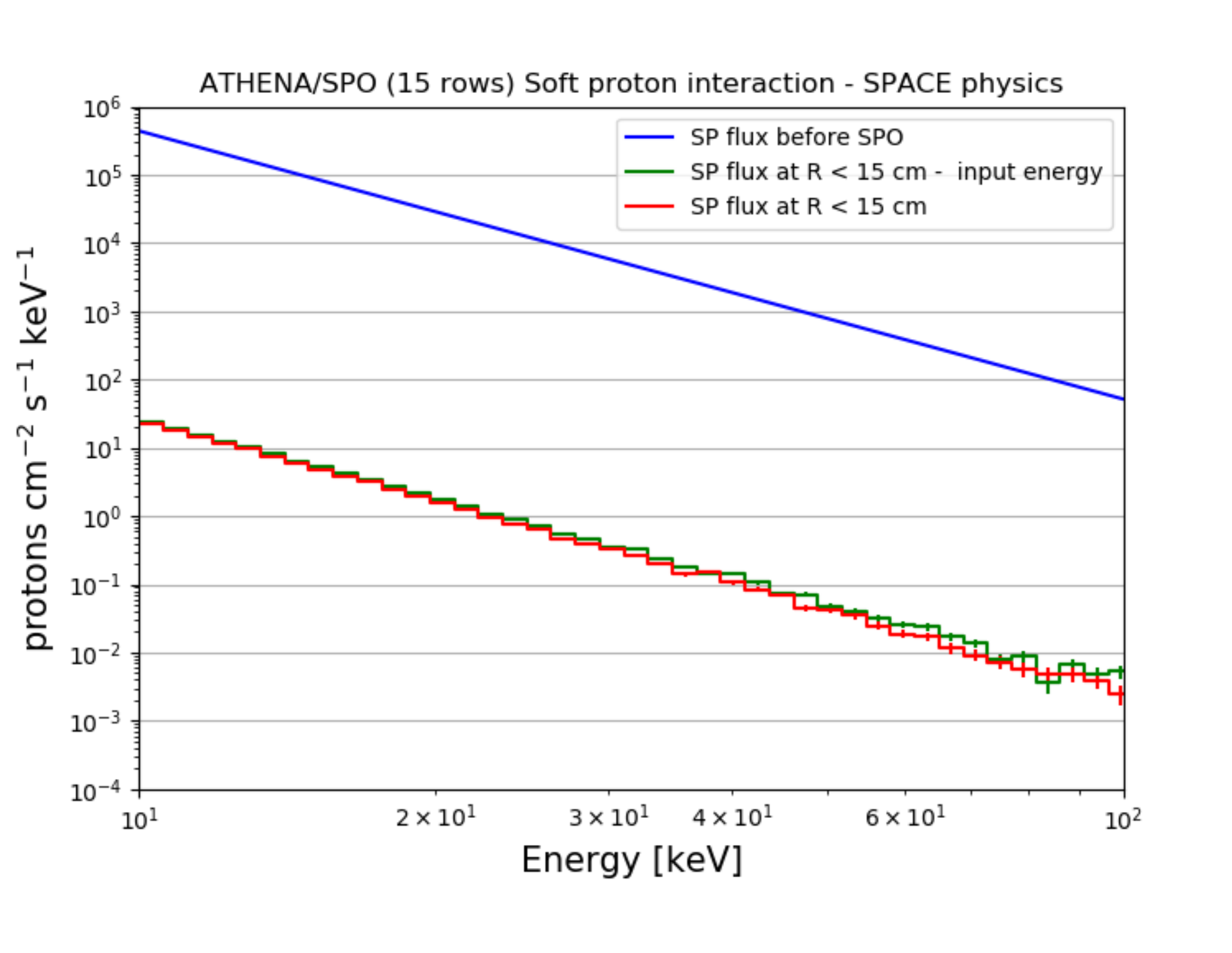}
   \caption 
   { \label{fig:fp}The energy distribution for protons reaching the focal plane, defined as a 15 cm radius region at SPO focal point, taking into account the energy after scattering (red line) and the input energy (green line) before entering the optics. The blue line shows as reference the input flux, integrated over $\pi$ sr, hitting the SPO entrance.}
   \end{figure*} 
\subsection{SPO mass model}
The SPO X-ray focusing technology \citep{2017SPIE10399E..0CC} is based on Silicon (Si) wafers cut by parallel grooves to create ribs. The ribbed Si plates are stacked together to form millions of pores, with an inner side coated by reflective metals. The Wolter-I configuration is realized in a mirror module composed by two parabolic and hyperbolic pore stacks, mounted in a common bracket. Modules are arranged in circular rows, for a total of 15 rows composing the full SPO at present design. The outer diameter of the 15th row is more than 2 m, with a focal length of 12 m.  
\\
Developing the Geant4 mass model, with hundreds of millions of volumes, while keeping a feasible CPU processing time, represented a challenge in itself and required the use of Geant4 optimization techniques along with a series of geometry simplifications (see \citet{fioretti_SPIE_2018} for a detailed description of the mass model). The resulting code (Fig. \ref{fig:spo}, left panel), converted into GDML\footnote{Geometry Description Markup Language, a specialized XML-based language designed for the description of Geant4 volumes.} format, was delivered to ESA as part of the official ATHENA mass model of the AREMBES project.

\subsection{Scattering physics}
Laboratory measurements \citep{2015ExA....39..343D} of the energy and angular distribution of low energy protons scattered by a sample of X-ray mirror shells, where compared in \citet{2017ExA....44..413F} to Geant4 simulations using two physics models to describe the interaction:
\begin{itemize}
\item the Remizovich solution \citep{1983RadEffect, 1980JETP...52..225R}, describing particles reflected by solids at grazing angles in terms of the Boltzmann transport equation, in the approximation of no energy losses (implemented in Geant4 as part of AREMBES activities);
\item the Geant4 single scattering (SS) model of the Coulomb scattering with the electron field of the nuclei, where each single interaction is computed contrary to the multiple scattering models (e.g. Urban or WentzelVI  models) that average the proton energy and angle over a larger number of interactions (see \citet{2017ExA....44..437I} and references therein for further details).
\end{itemize}
The cited measurements did not covered the low scattering angle ($<1^{\circ}$) and low energy ($<250$ keV) regimes. These are the regimes that are expected to affect the most the proton focusing effect in X-ray telescopes. Despite this caveat (see Sec. \ref{sec:con}), both Remizovich and SS models are able to describe the general behavior of the measured scattering efficiency \citep{2017ExA....44..413F}.
Since the SS model also foresees proton energy losses, although lower than the measured ones, and its Geant4 implementation is better optimized in terms of CPU performance, we rely on it in the simulation of soft proton scattering with the SPO mass model. The AREMBES \textit{SPACE} physics list \citep{arembes_wp3}, optimized for space applications, includes the SS model for protons $<$1 MeV and it used throughout the presented work.
\begin{figure*} 
    \includegraphics[trim={5cm 8cm 3cm 9cm},clip,width=0.55\textwidth]{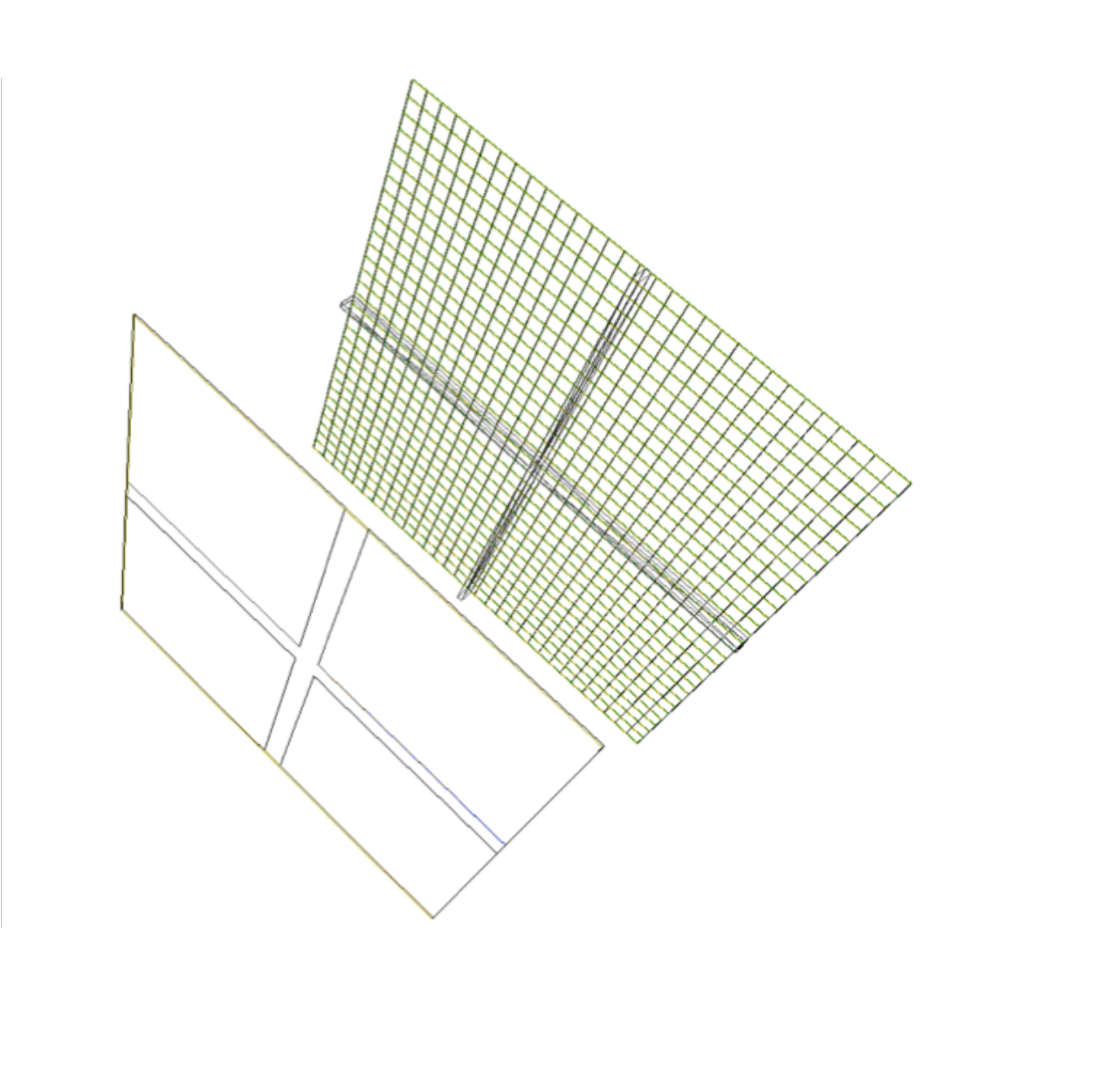}
   \includegraphics[trim={5cm 8cm 6cm 9cm},clip,width=0.4\textwidth]{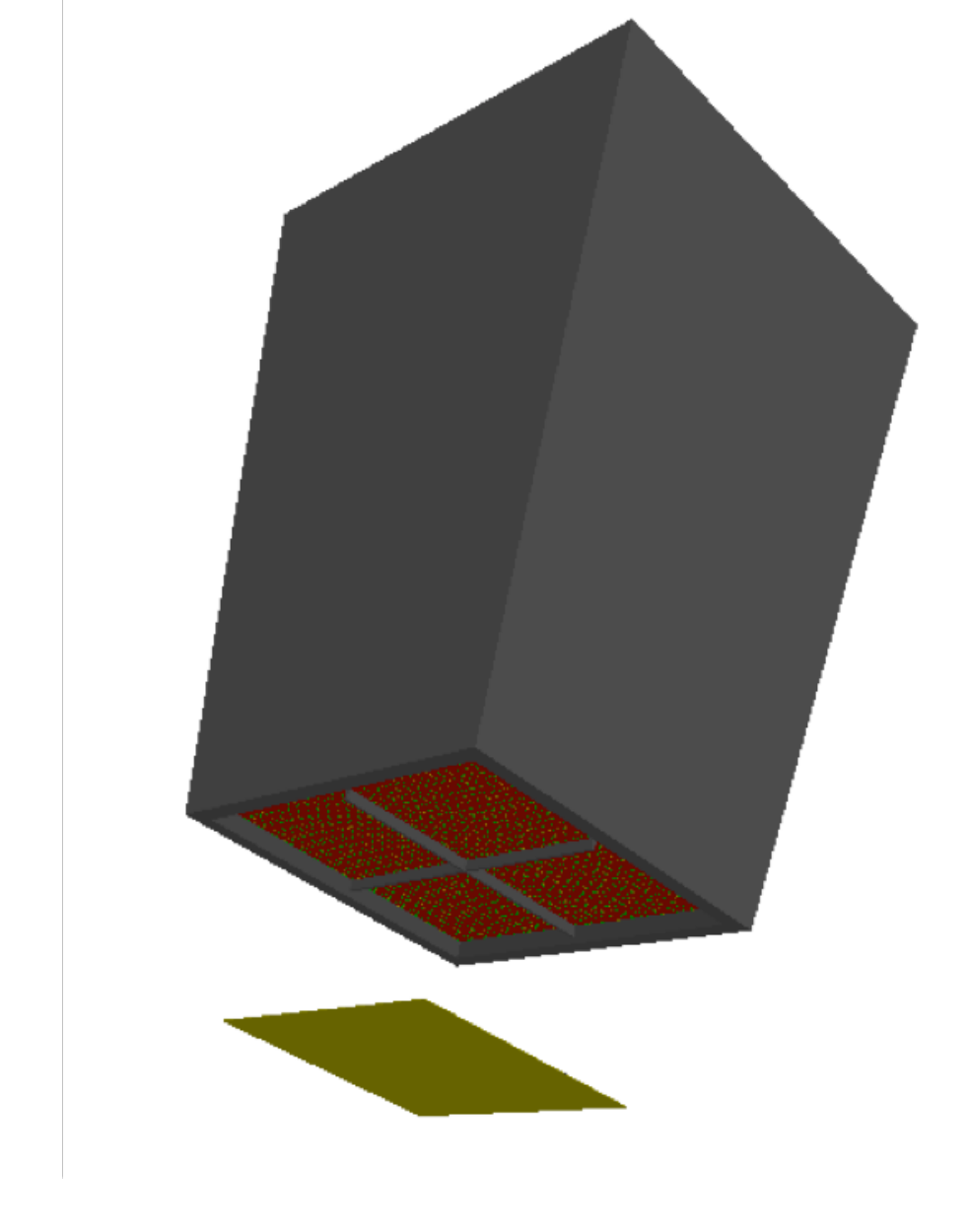}
   \caption 
   { \label{fig:wfi}\textit{Left panel:} wire-framed visualization of the WFI (bottom-left) and the filter wheel optical filter (top-right) mass model. \textit{Right panel:} Aluminum-based baffle mass model, shown in grey, placed at the base of the filter wheel and on top of the WFI, shown here as squared slab for visualization purposes.}
   \end{figure*} 
\subsection{Residual flux on ATHENA focal plane}
The input proton flux is simulated by means of 15 independent runs, one for each row, from an annular surface matching the row entrance area within a cone of $5^{\circ}$ half-angle aperture. Above this angle the effective area to proton scattering rapidly falls \citep{fioretti_SPIE_2018}. The simulated angles follow a cosine-law distribution to ensure the same flux from each incident direction \citep{fioretti_norm}. The binned spatial distribution of the protons exiting the SPO is shown in Fig. \ref{fig:spo} (right panel): a higher number of protons exits the outer rows, because of the shorter length of the mirror modules.
We define as focal plane a circular region of 15 cm radius at the 12 m focal distance, starting from the intersection between the paraboloid and the hyperboloid sections. Despite being much larger than the actual WFI ($<10$ cm) and X-IFU ($<$1 cm, \citet{2016SPIE.9905E..2FB}) radial size, this selection includes the filter wheel baffle on top the WFI, with a 26 cm total side (see Sec. \ref{sec:wfi}), and the X-IFU cryostat baffle, with a 21.9 cm diameter. Protons within the focal plane selection will be used as input for the simulation of the WFI X-ray background in Sec. \ref{sec:res}. The residual proton flux reaching the focal plane is shown in Fig. \ref{fig:fp} (red line). The input proton flux, in blue, is integrated over the hemisphere solid angle including the flux cosine dependence:
\begin{equation}
 \Omega = \int_{0}^{2\pi}\int_{0}^{\rm \pi/2}\rm cos\rm \theta\rm sin\theta\rm d\rm \theta\rm d\phi = \pi\:\;\rm sr\:,
\end{equation}
where $\theta$ starts from the telescopes axis, to take into account all the protons entering the mirror pores. If a 15 row SPO configuration is used, a flux of 20 protons cm$^{-2}$ s$^{-1}$ keV$^{-1}$ at 10 keV reaches the focal plane, $2\times10^{4}$ times lower than the primary flux entering the mirror, for the same energy. Since the SS model foresees a energy loss of few percent with respect to the input one, the energy distribution of the primary energy for the protons reaching the focal plane is almost unchanged.

\section{Soft proton induced WFI background}\label{sec:wfi}
The WFI non X-ray background (NXB) contribution given by charged particles transmitted through the mirrors must be $<10\%$ of the non focused component requirement, currently defined as $5\times10^{-3}$ counts cm$^{-2}$ s$^{-1}$ keV$^{-1}$ in the 2 -- 7 keV energy range \citep{2017SPIE10397E..0VM}. Assuming the focused NXB being dominated by protons, we require a soft proton induced background $<5\times10^{-4}$ counts cm$^{-2}$ s$^{-1}$ keV$^{-1}$ in the 2 -- 7 keV energy range.
The product of the SPO Geant4 simulation, a proton list with associated energies and directions, is used as input in the simulation of the interaction with the WFI assembly, comprising the detector itself, the optical stray-light baffle and the optical filters. The WFI is in fact also sensitive to UV and optical photons entering the field of view which can degrade the spectral performance of the detector. X-ray ultra-thin optical filters, required to block this contamination (see \citet{2015SPIE.9601E..09B} for a review), have the side effect of slowing down soft protons reaching the detector aperture, and their configuration impacts on the residual background count rate. The WFI design foresees two sets of optical filters, the fixed on-chip optical blocking filters (OBF from now on) and the optical blocking filter accommodated in the filter wheel assembly (FW from now on).
The filter and calibration wheel \citep{2016SPIE.9905E..68R}, placed at 10 cm from the focal plane, can have two positions during sky observations: open aperture and optical blocking filter.
In the first position, only the OBF is present, while in the second both the OBF and the additional FW filters are placed in the field of view. 
The WFI Geant4 mass model (Fig. \ref{fig:wfi}) comprises:
\begin{itemize}
\item the wide field detector itself, composed by four pixelated quadrants each with $512\times512$, $130\times130$ $\mu m^{2}$ side pixels, with a thickness of the Si sensor of 450 $\mu m$ (bottom-left of Fig. \ref{fig:wfi}, left panel);
\item the OBF, placed on top the four quadrants and composed, from top to bottom, by 90 nm of Aluminum (Al), 30 nm of Silicon Nitride (Si$_{3}$N$_{4}$) and 20 nm of Silicon Oxide (SiO$_{2}$);
\item the squared $17\times17$ cm$^{2}$ FW  (top-right of Fig. \ref{fig:wfi}, left panel), composed from top to bottom by 30 nm of Al and 150 nm of Kapton (polyimide) as blocking filters, a supporting mesh of 0.2 mm thick stainless steel coated by 5 $\mu m$ of Gold (Au), and the inner bars of the Al frame, 6 mm thick;
\item the Al squared baffle (Fig. \ref{fig:wfi}, right panel), 1 cm thick, on top of the filter wheel is also simulated to evaluate the potential impact of protons back-scattered by the baffle walls and reflected to the WFI, especially in case of a magnetic diverter in place.
\end{itemize}
The error bars in the following results are 1$\sigma$ statistical Poisson fluctuations given by the number of simulated particles.

\subsection{Residual particle flux on WFI}\label{sec:pres}
\begin{figure*} 
    \includegraphics[trim={3cm 2.5cm 4.5cm 2cm},clip,width=0.49\textwidth]{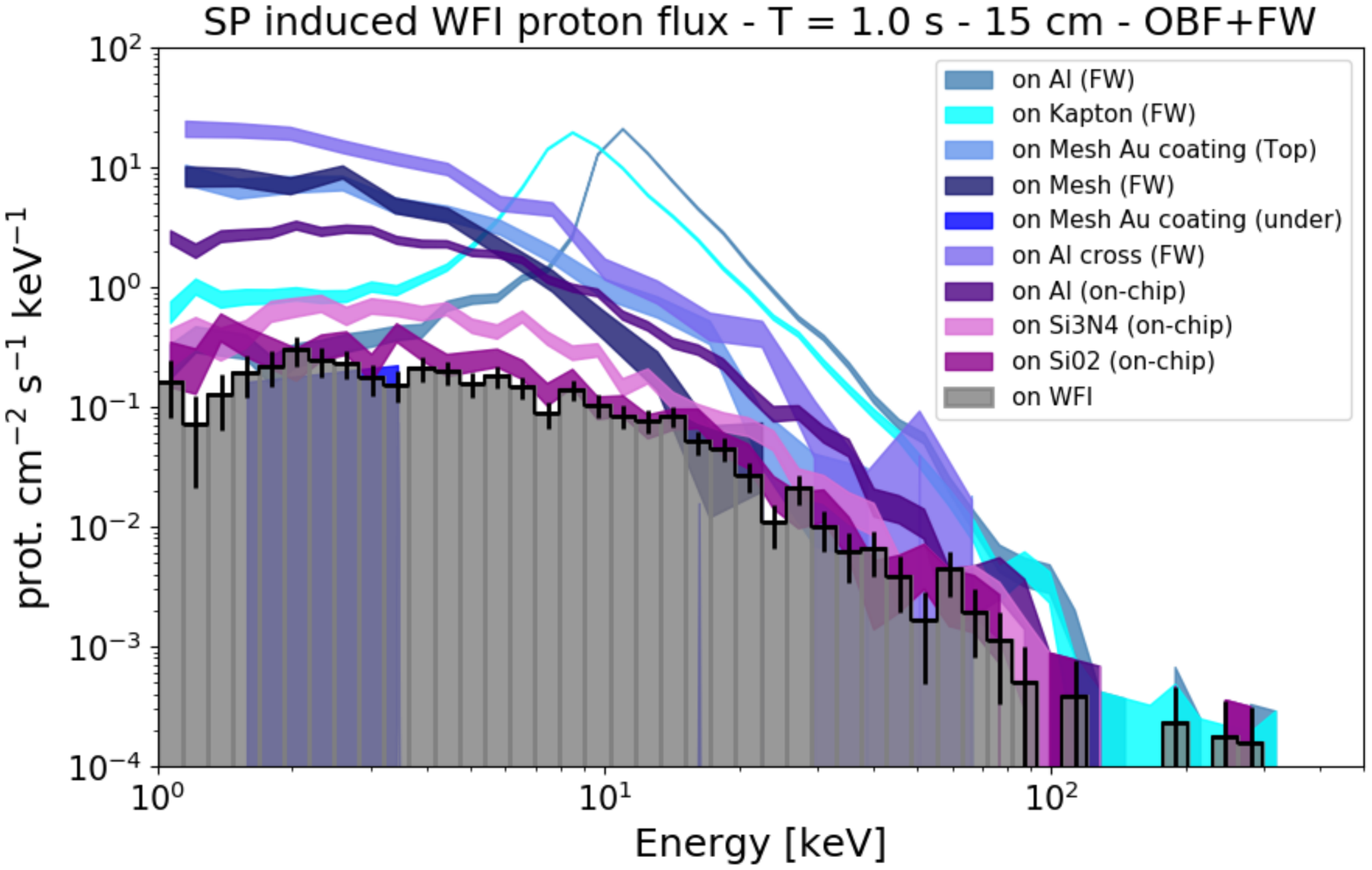}
   \includegraphics[trim={2.5cm 2.5cm 4.5cm 2cm},clip,width=0.5\textwidth]{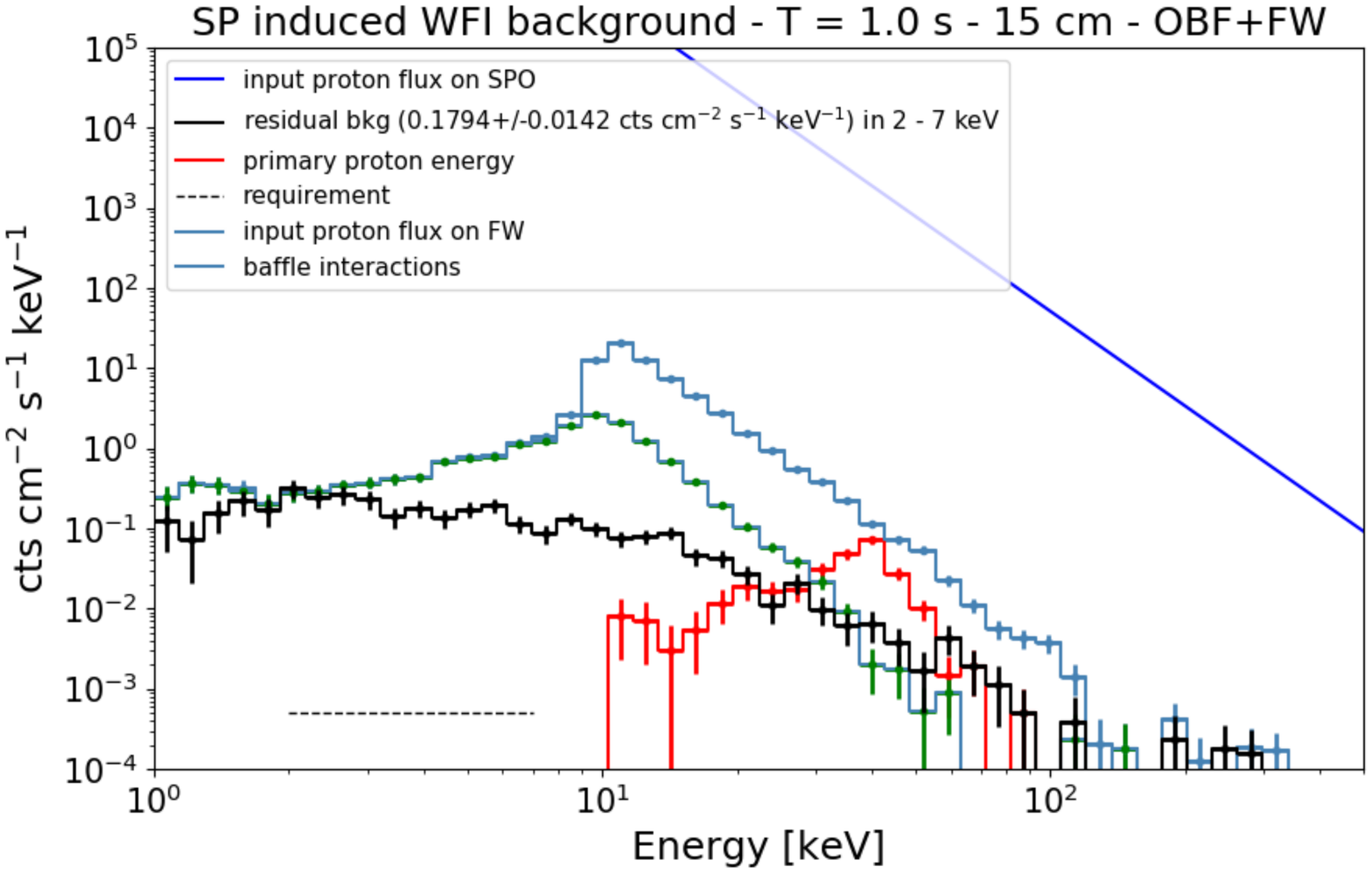}
\\
   \includegraphics[trim={3cm 2.5cm 4.5cm 2cm},clip,width=0.49\textwidth]{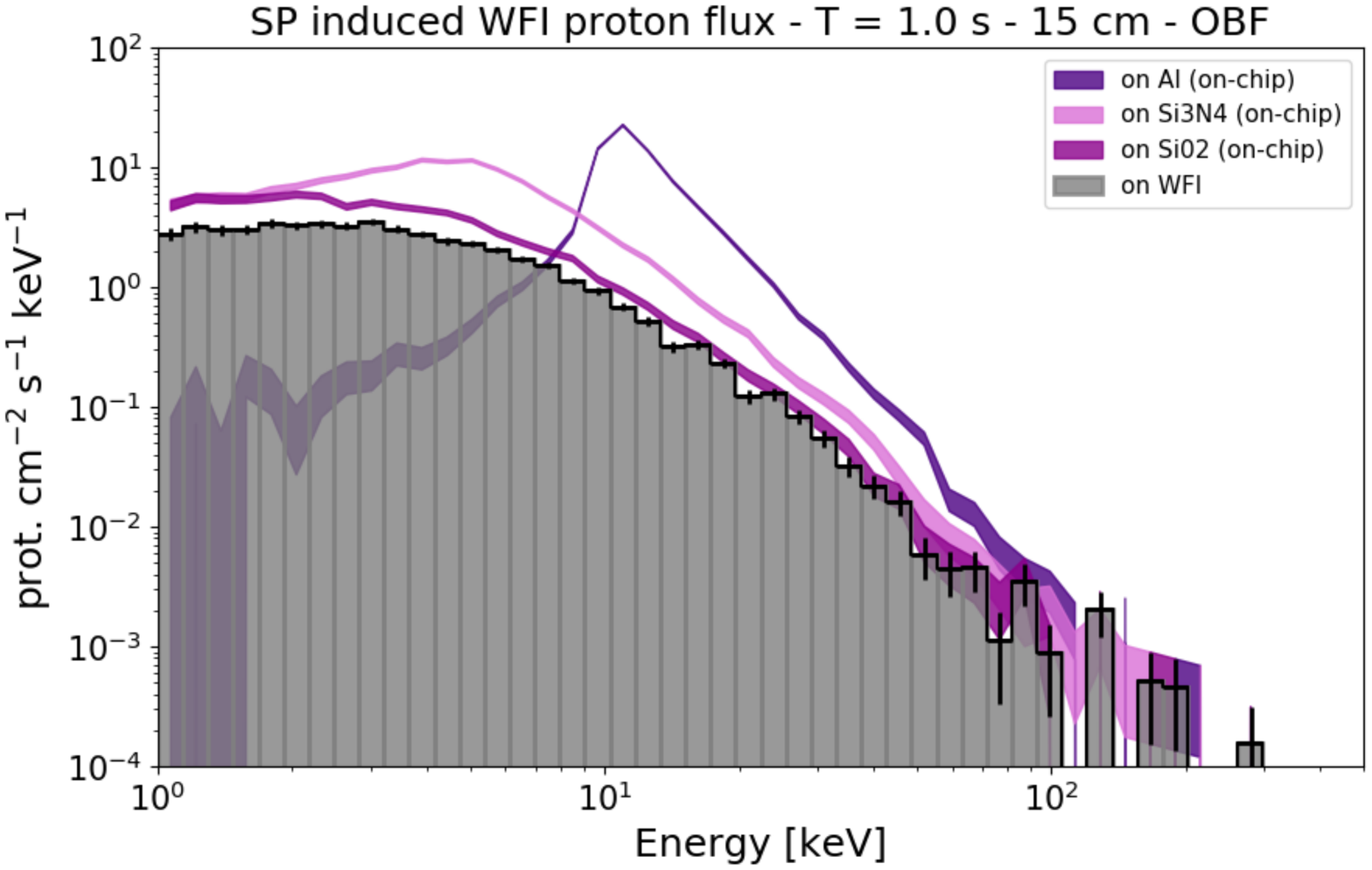}
   \includegraphics[trim={2.5cm 2.5cm 4.5cm 2cm},clip,width=0.5\textwidth]{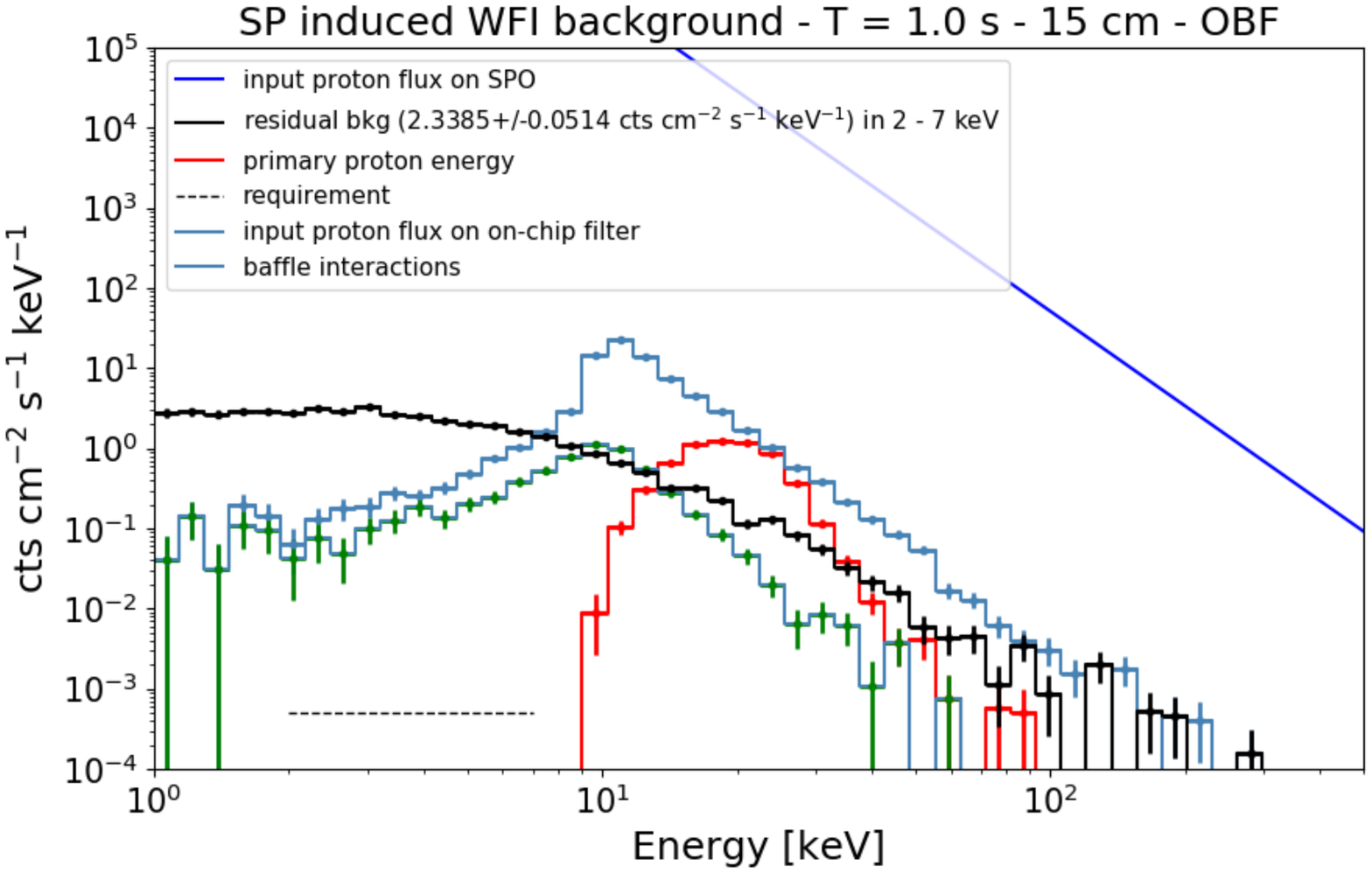}
\\
   \caption 
   { \label{fig:bkg} \textit{Left panels:} proton flux hitting each layer of the OBF+FW (top panel) and the OBF (bottom panel) set of stopping layers. \textit{Right panels:} The input proton flux (dark blue line), the total proton flux on the first optical filter layer (light blue line) and its component preceded by baffle interactions (green line), the background count rate seen by the WFI (black line) and the correspondent primary energy of protons (red line) inducing the background counts in 2 -- 7 keV for a configuration including the FW filters (top panel) and with just the OBF (bottom panel). The dashed line refers to the requirement on the focused NXB component.}
   \end{figure*}

Soft protons interact with each layer or structure placed in the field of view, depositing a portion of its energy. Fig. \ref{fig:bkg} (left panels) shows from top to bottom the energy spectrum of protons hitting these layers, with the first one being the FW Al filter in the filter wheel assembly or the Al filter in the OBF, depending on the filter wheel configuration. Since the simulated lowest energy of the input population is 10 keV, protons hitting the first layer show a power-law shape above 10 keV plus a low energy tail. The latter is caused by the combination of multiple, but rare, reflections on the mirror surface and scatterings with the baffle walls. See \citet{fioretti_SPIE_2018} for a detailed study on the energy and angular distribution of protons at the mirror exit. 
\\
If the FW filter is moved in front of the detector, protons hitting each layer, w.r.t. the first Al layer, are 97\% on kapton, 12\% on on-chip Al, 3\% on Si$_{3}$N$_{4}$ and 2\% on SiO$_{2}$. Less than 1\% of protons hit the supporting mesh, as expected given that the mesh only covers the 3.7\% of the filter surface.
Since protons interact with the FW at 10 cm of distance from WFI, the combined effect of the FW stopping power and the induced angular spread on the proton distribution causes a factor 10 reduction in protons hitting the OBF. 
\\
Because of the lower stopping power, the total number of protons reaching the WFI sensitive pixels is 10 times higher if only the on-chip filters are placed in the field of view. 
Since the stopping power for protons reaches its maximum in the 50 -- 100 keV range for the materials under study, and decreases at lower energies, the effect of the filter interaction on the proton energy is not a linear shift towards lower energies, and proton hitting the WFI keep a weak trace of the input power-law spectral model.
\\ 
We also searched for the presence of electrons generated by proton induced ionization in the filters and in the X-ray baffle but its impact can be neglected since they contribute $<0.1\%$ of the particles hitting the WFI.
\subsection{Vignetting}
The term vignetting usually refers to the reduction of the collective area of the mirrors as a function of the off-axis angle of incoming X-ray photons. It leads to a decrease in the counts density towards the border of the detection plane. A similar effect was observed by XMM-Newton for soft protons. The EPIC MOS observed a radial distribution of scattered soft protons \citep{kun08} with a factor 2 reduction from the center to the edge. 
\\
In Fig. \ref{fig:vign} (left panel) we investigate the presence of soft proton vignetting for the WFI Large Detector Array. For each quadrant, the count radial distribution, in the 2 -- 7 keV, is computed from the inner edge and then binned. 
The total number of counts in each annulus is then divided by its area to obtain the count surface density. Both distributions, with and without the FW OBF, are fitted by a constant and the reduced $\chi^{2}$ of the best fit is reported in figure. If only the on-chip OBF is present, a small reduction of $\sim25\%$ is found the center to the edge. The same weak vignetting is reported in the surface density of the protons reaching the focal plane \citep{fioretti_SPIE_2018}. 
No vignetting is found if the FW filters are present because the scattering with its filters spreads the proton angular distribution.
\subsection{Residual X-ray background spectrum}\label{sec:res}

Each count is computed by summing all the energy deposits within the same pixel and applying an energy threshold of 0.2 keV. We find that protons interacting with the WFI cause single pixel counts, without stripes or clusters of pixels usually observed from the interaction of cosmic galactic rays. Same single pixel events are measured in XMM-Newton soft proton flare detections.
The X-ray flux (black line of Fig. \ref{fig:bkg}, right panel) is
constant up to 5 keV and starts to decrease at higher energies.
The X-ray spectrum of protons hitting the first layer of the OBF or FW is plotted in light blue. It is clear that optical filters lower up to 100 times the soft proton induced background rate and that their specific configuration modifies the WFI background with the same degree as for the different input plasma regimes. The green line refers to the protons that interact first with the baffle and then hit the first optical layer, contributing for most of the $<10$ keV events. A fraction of the protons scattering within the baffle walls is redirected towards the WFI. These particles lose a significant amount of energy in the process and increasing the low energy proton tail. The baffle contribution to the soft proton flux on the focal plane is not negligible, for an input proton population with energies $>10$ keV.
 \begin{figure*}
   \includegraphics[trim={1cm 2cm 2cm 3.cm},clip,width=0.53\textwidth]{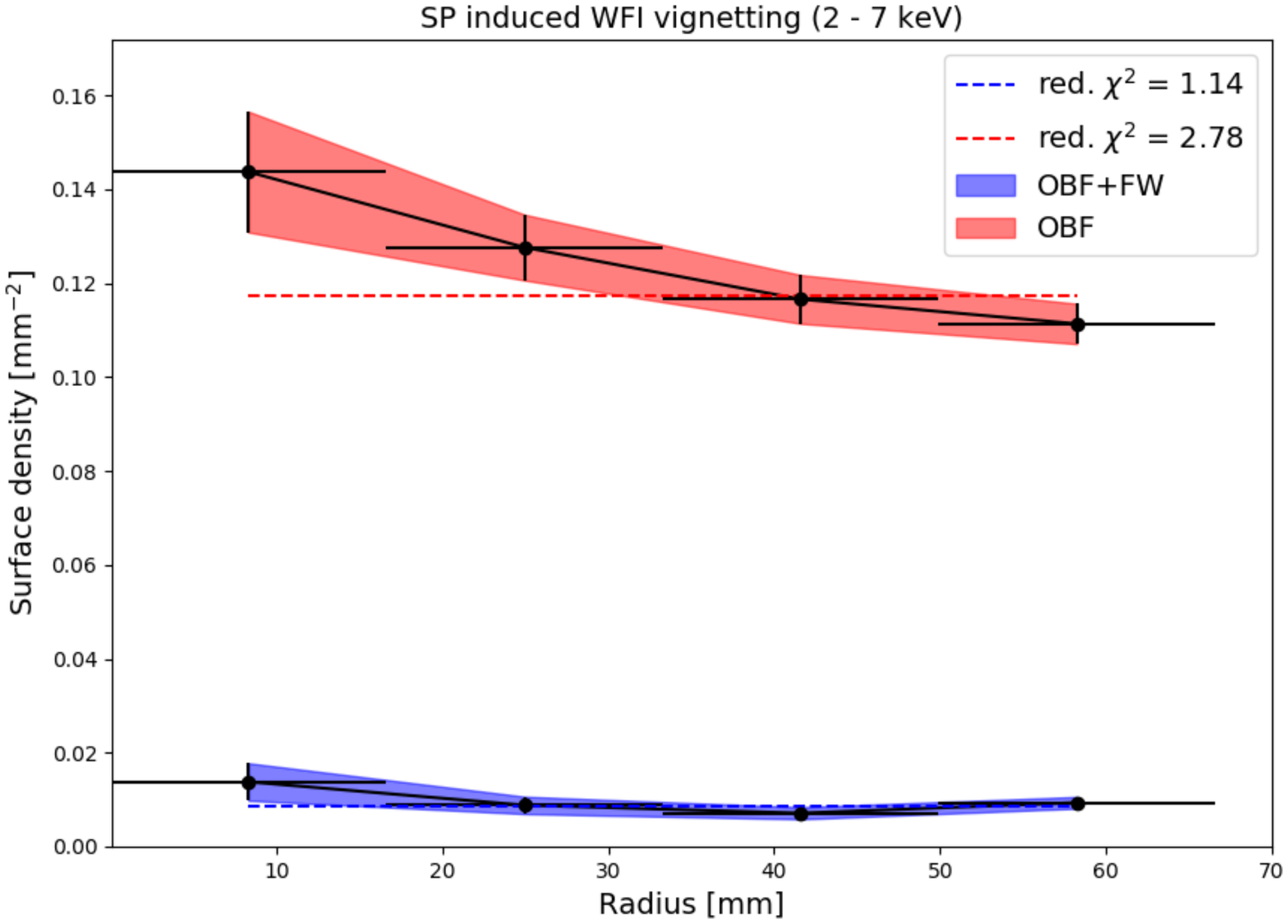}
   \includegraphics[trim={2cm 9cm 2.cm 8cm},clip,width=0.47\textwidth]{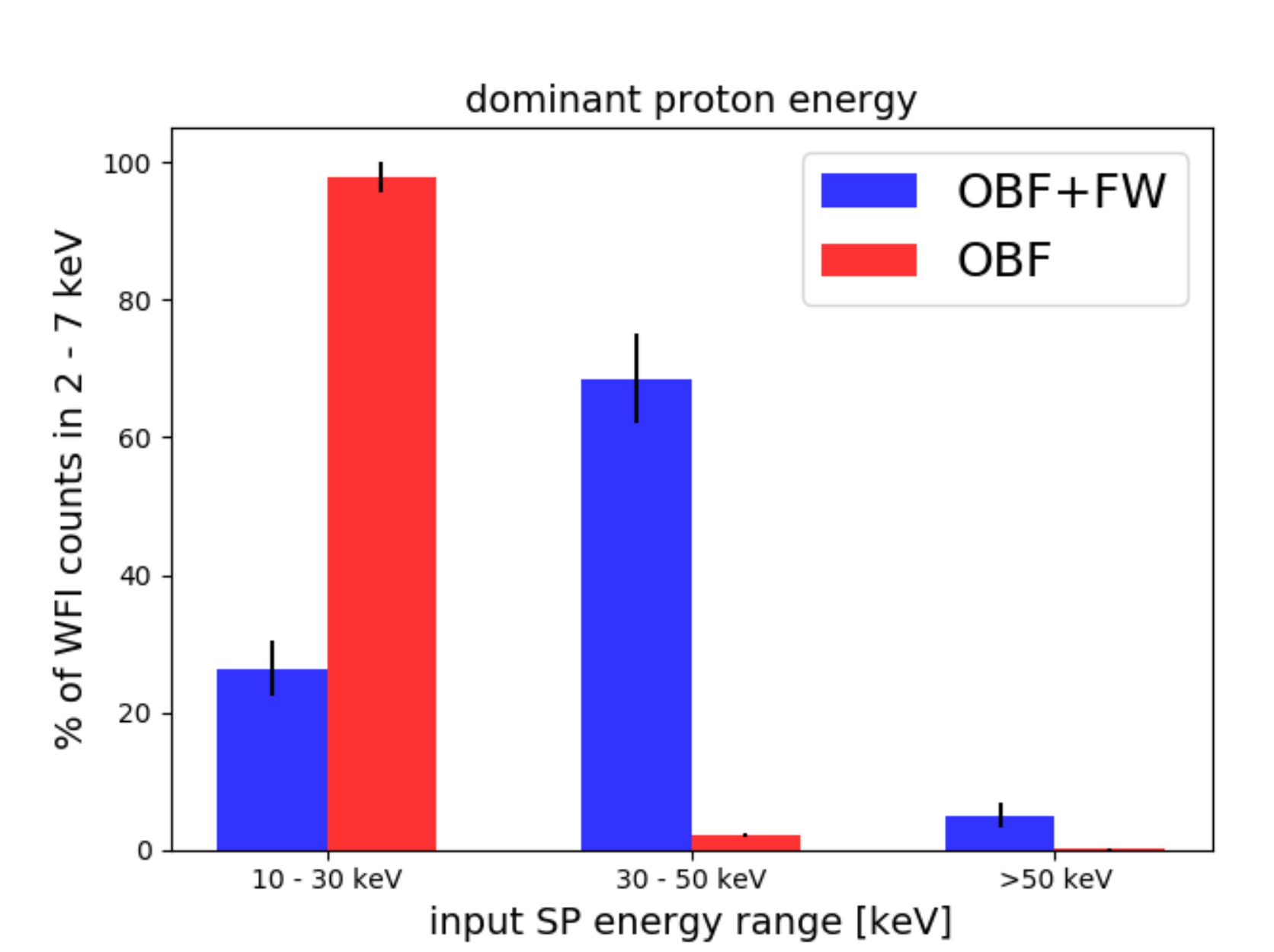}
   \caption 
   { \label{fig:vign}\textit{Left panel:} the radial distribution of WFI background counts plotted as surface density with (red curve) and without (blue curve) the filter wheel optical filter. Both curves are fitted by a constant and the reduced $\chi^{2}$ resulting from the best fit is shown in legend. \textit{Right panel:} percentage of WFI background counts, in the 2 -- 7 keV energy range, generated by three selected input soft proton energy bands: $<30$ keV, 30 -- 50 keV, and $>50$ keV for the two filter configurations.}
   \end{figure*}
  \begin{figure*}
   \includegraphics[trim={5cm 1cm 5cm 2.5cm},clip,width=0.5\textwidth]{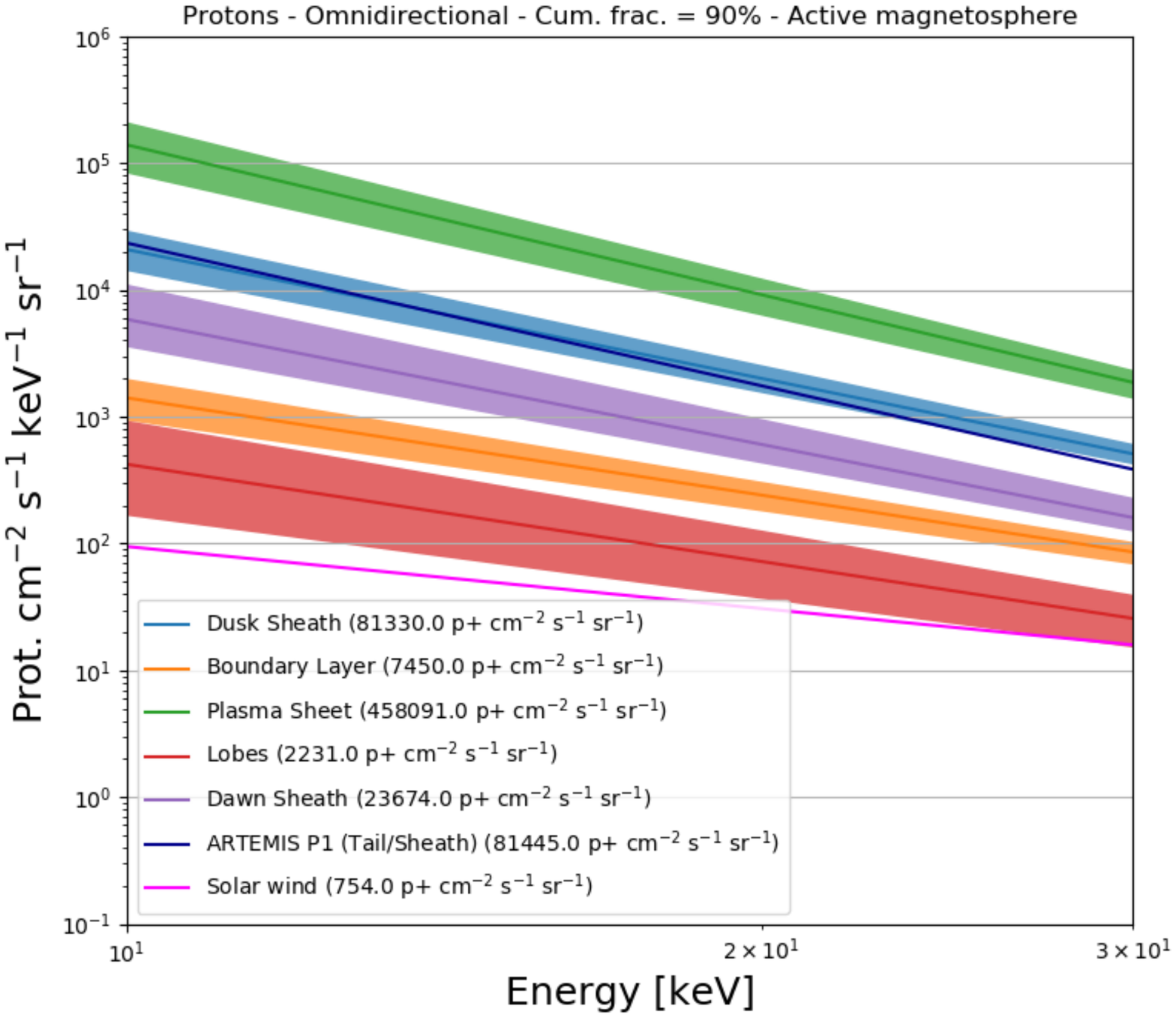}
   \includegraphics[trim={1cm 6cm 2cm 6.cm},clip,width=0.49\textwidth]{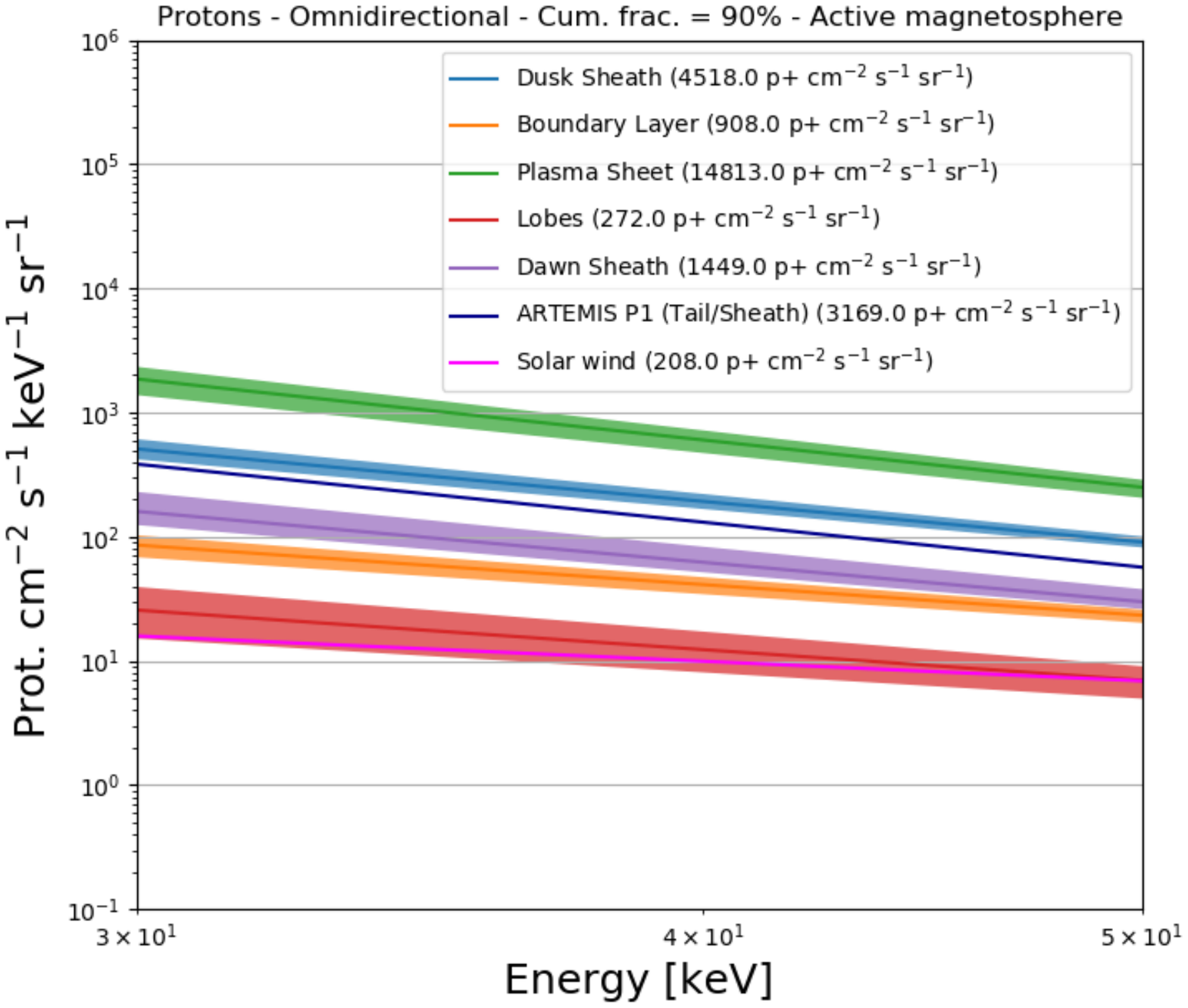}
   \caption 
   { \label{fig:zoom} Input proton models for each magnetotail plasma regime and the solar wind in the 10 -- 30 keV (left panel) and the 30 -- 50 keV (right panel) energy range. Values in brackets report the corresponding energy integrated fluxes.}
   \end{figure*} 
\\
\begin{table*}
\begin{center}
\begin{tabular}{c|c|c|c|c}
\multicolumn{5}{c}{\textsc{Soft proton WFI background induced by focused NXB}}\\
\hline
\hline
\multirow{2}{*}{Plasma regime} & \multirow{2}{*}{OBF$+$FW}& Red. factor &  \multirow{2}{*}{OBF}  &Red. factor \\
 & & [30 - 50 keV] & & [10 - 30 keV] \\
\hline
\multirow{2}{*}{Plasma Sheet}  &  \multirow{2}{*}{0.1794 ($359\times$)}  & \multirow{2}{*}{1}& \multirow{2}{*}{2.3385 ($4677\times$)} &\multirow{2}{*}{1}  \\
& & & &\\
\multirow{2}{*}{Dusk Sheath}  & \multirow{2}{*}{0.0547 ($109\times$)}& \multirow{2}{*}{0.305} & \multirow{2}{*}{0.4116 ($823\times$)} & \multirow{2}{*}{0.176}\\
& & & &\\
\multirow{2}{*}{Tail/Sheath}  & \multirow{2}{*}{0.0384 ($77\times$)}  & \multirow{2}{*}{0.214}& \multirow{2}{*}{0.4163 ($833\times$)}  & \multirow{2}{*}{0.178}\\
& & & &\\
\multirow{2}{*}{Dawn Sheath} & \multirow{2}{*}{0.0176 ($35\times$)}  & \multirow{2}{*}{0.098} & \multirow{2}{*}{0.1216 ($243\times$)} & \multirow{2}{*}{0.052}\\
& & & &\\
\multirow{2}{*}{Boundary layer}  & \multirow{2}{*}{0.0109 ($22\times$)}& \multirow{2}{*}{0.061}    & \multirow{2}{*}{0.0374 ($75\times$)}& \multirow{2}{*}{0.016}\\
& & & &\\
\multirow{2}{*}{Lobes}  &  \multirow{2}{*}{0.0032 ($6\times$)}& \multirow{2}{*}{0.018}& \multirow{2}{*}{0.0117 ($23\times$)}& \multirow{2}{*}{0.005} \\
& & & &\\
\multirow{2}{*}{Solar wind} &  \multirow{2}{*}{0.0025 ($5\times$)} & \multirow{2}{*}{0.014} & \multirow{2}{*}{0.0047 ($9\times$)}& \multirow{2}{*}{0.002}\\
& & & &\\
\hline
\end{tabular}
\end{center}
\caption{\label{tab:comp}Soft proton induced WFI background in counts cm$^{-2}$ s$^{-1}$ keV$^{-1}$ in the 2 -- 7 keV energy range for each plasma regime populating the L2 environment and for both filter configurations (OBF$+$FW and OBF). The result of the comparison with the focused NXB requirement is shown within brackets.}
\end{table*}  
The Geant4 simulation allows to record the input energy of the protons that generate background counts on the WFI in the 2 -- 7 keV band, after scattering with the SPO and crossing the optical filters. The spectral distribution of the primary energy of protons generating counts in the 2 -- 7 keV energy range is plotted in red in Fig. \ref{fig:bkg} (right panels) and outlines two different results depending on the filter configuration. We divide the input protons in three ranges (10 -- 30 keV, 30 -- 50 keV, $>50$ keV) and their respective percentage contributing to the WFI counts is shown in Fig. \ref{fig:vign} (right panel). If only the OBF is in place, $98\%$ of protons have energies in the 10 -- 30 keV band, decreasing rapidly below 10 keV. On the contrary, if the FW is present, the proton energy peaks at 40 keV, with 69\% protons in the 30 -- 50 keV energy range, but an additional contribution of $26\%$ is also present at lower energies, mainly induced by interactions with the baffle walls. 
The uncertainty in the energy integrated plasma sheet flux, given by 2 times the fit r.m.s., is 31\% and 15\% for the two dominant energy ranges of 10 -- 30 keV and 30 -- 50 keV. We can evaluate the consequent systematic uncertainty on the background flux and propagate it with the systematic error given by the simulation. The residual soft proton induced background, for the proton population expected in the plasma sheet magnetotail regime, is $0.18\pm0.01$ counts cm$^{-2}$ s$^{-1}$ keV$^{-1}$ (OBF$+$FW) and $2.3\pm0.05$ counts cm$^{-2}$ s$^{-1}$ keV$^{-1}$ (OBF) in the 2 -- 7 keV energy range, well above the requirement. 
\subsection{Background flux in L2}
As mentioned in Sec. \ref{sec:l2}, the plasma regimes in L2 are characterized by different power-law slopes, hence each of them should require a dedicated simulation.
However, since only a small portion of the proton spectrum contributes to the WFI background in the 2 -- 7 keV band, the models are, within the uncertainties, very similar, if selected in the dominant 10--30 keV and 30 -- 50 keV bands (Fig. \ref{fig:zoom}). Therefore, we can compute the energy integrated flux for each input model and use the reduction factor with respect to the plasma sheet integral flux to propagate the simulated soft proton induced background to all the regimes populating the magnetotail. The averaged uncertainty of all plasma regimes is 46\% in the 10 -- 30 keV interval and 24\% in the 30 -- 50 keV interval. We assume the same error in the propagated background fluxes, for the two filter wheel configurations (Tab. \ref{tab:comp}).
\\
The resulting soft proton induced background, for a cumulative fraction of 90\% and an active magnetosphere, is always above the requirement, including the minimum value of $2.5\pm0.5\times10^{-3}$ counts cm$^{-2}$ s$^{-1}$ keV$^{-1}$ (OBF$+$FW) in the solar wind regime. 
The background flux derived from the ARTEMIS survey of the magnetotail/magnetosheath region is well consistent with independent Geant4 simulations carried on by ESA \citep{2017DUSANreport}, while we reproduce at the order of magnitude level the results of \citet{lotti_2018_submitted}. This is expected since the latter approximates both the physics of interaction with the optics, using the Remizovich model in the approximation of no energy losses, and the interaction with the optical filters, not including the distance of the filter wheel.
\section{Proton diverter shielding}\label{sec:div}
 \begin{figure*} 
   \includegraphics[trim={1cm 7cm 3cm 6cm},clip,width=0.52\textwidth]{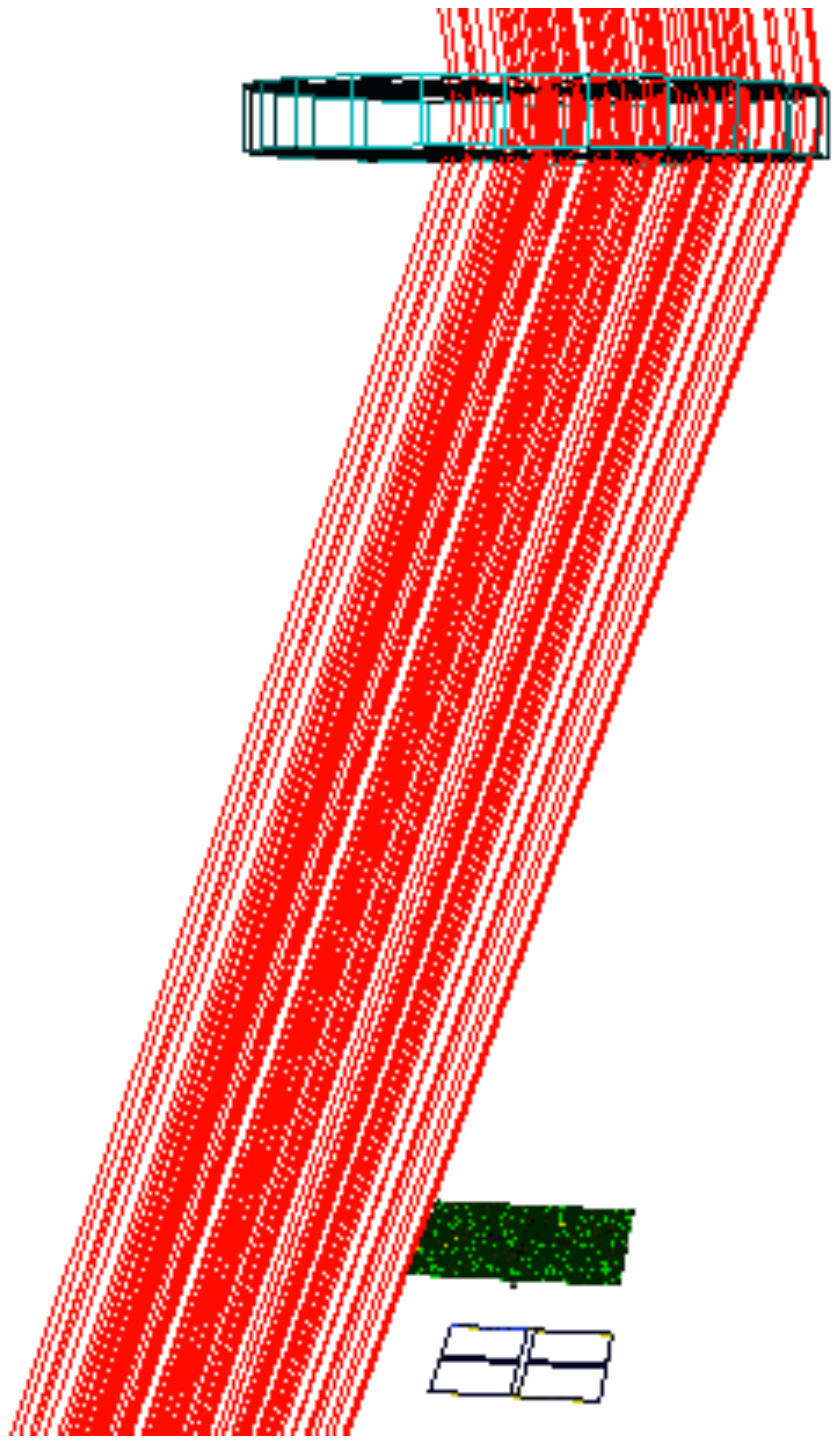}
   \includegraphics[trim={1cm 6cm 3cm 7.3cm},clip,width=0.48\textwidth]{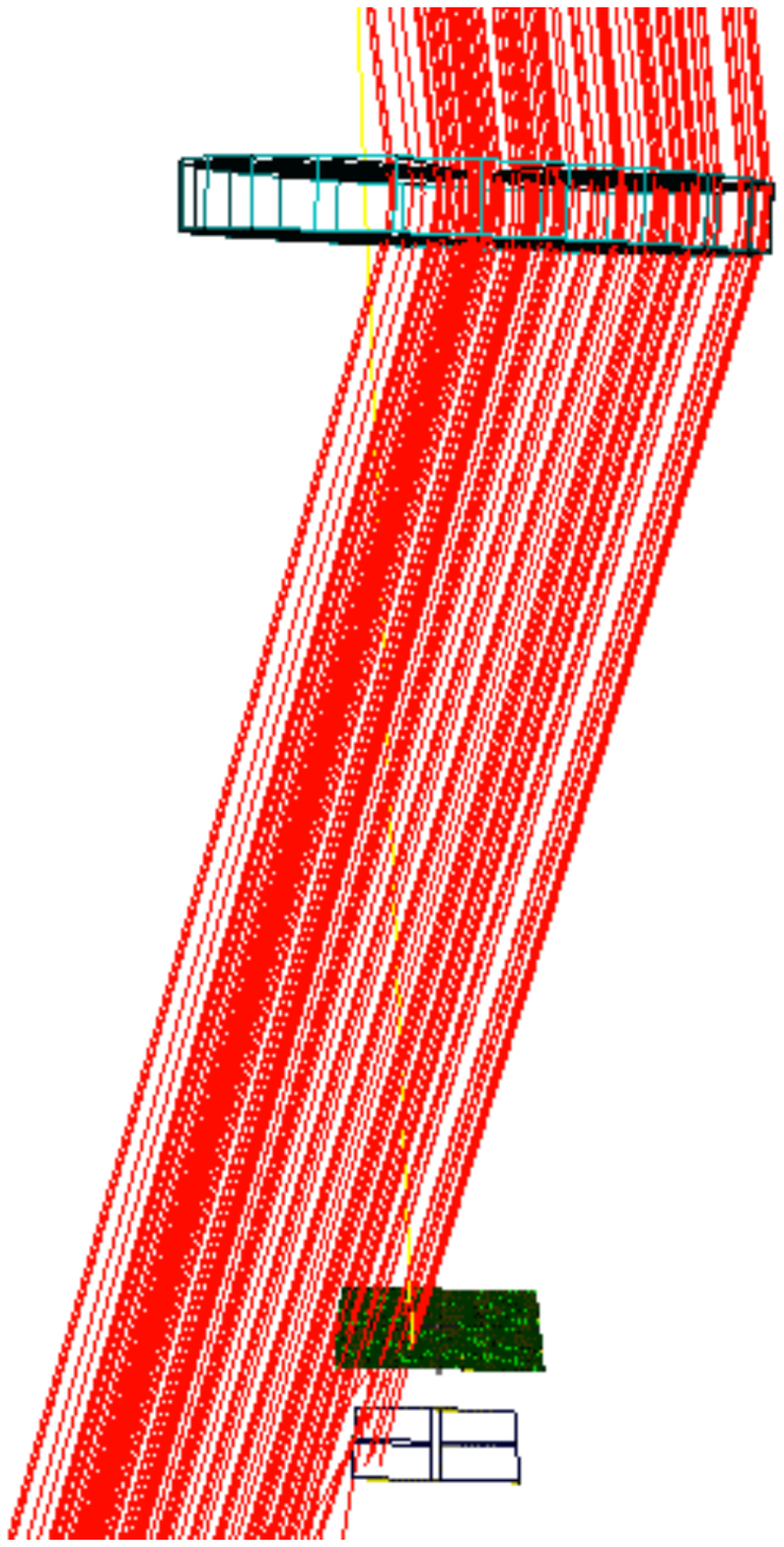}
   \caption 
   { \label{fig:MD_test} Testing the Geant4 implementation of the diverter field by simulating 76 keV (left panel) and 100 (right panel) keV protons as a parallel beam from the maximum incident angle. Red tracks represent the proton trajectory, the top disc the diverter inner region, and the bottom volumes the WFI plus the FW OBF on top of if. The X-ray baffle is not present to better visualize the proton trajectory.}
   \end{figure*} 
Contrary to the magnetic diverter proposed for the Simbol-X mission \citep{2008SPIE.7011E..2YS}, where permanent magnets were aligned on the X-ray optics spider structure, the current baseline design for ATHENA foresees diverters for both instruments to be placed near the science instrument module. The magnet arrangement will reproduce an Halbach cylinder \citep{esa_diverter}, where the continuous rotation of the magnetization generates an intense field within the cylinder and no field outside. Protons crossing the central region of the diverter would be deflected outside the WFI field of view. For an ideal Halbach cylinder, the field density is uniform and protons would be deflected towards the same direction. In the Geant4 simulation of the diverter field and its interaction with the soft proton beam, we use the WFI diverter configuration derived by analytical computations \citep{lotti_2018_submitted} in order to deflect all protons with energy below 76 keV. Below this threshold energy, 99\% of the residual protons are required to be deflected from the WFI detector. The field density of 0.38 T, with a cylinder height of 5 cm and an inner diameter of 45.6 cm, is computed for the worst case of deflecting protons entering the diverter with the maximum incidence angle and, i.e. protons exiting at the SPO edge and reaching the diverter at the opposite side. The diverter distance is 1 m from the WFI.
\begin{figure*}
   \includegraphics[trim={1cm 3cm 3cm 4cm},clip,width=0.49\textwidth]{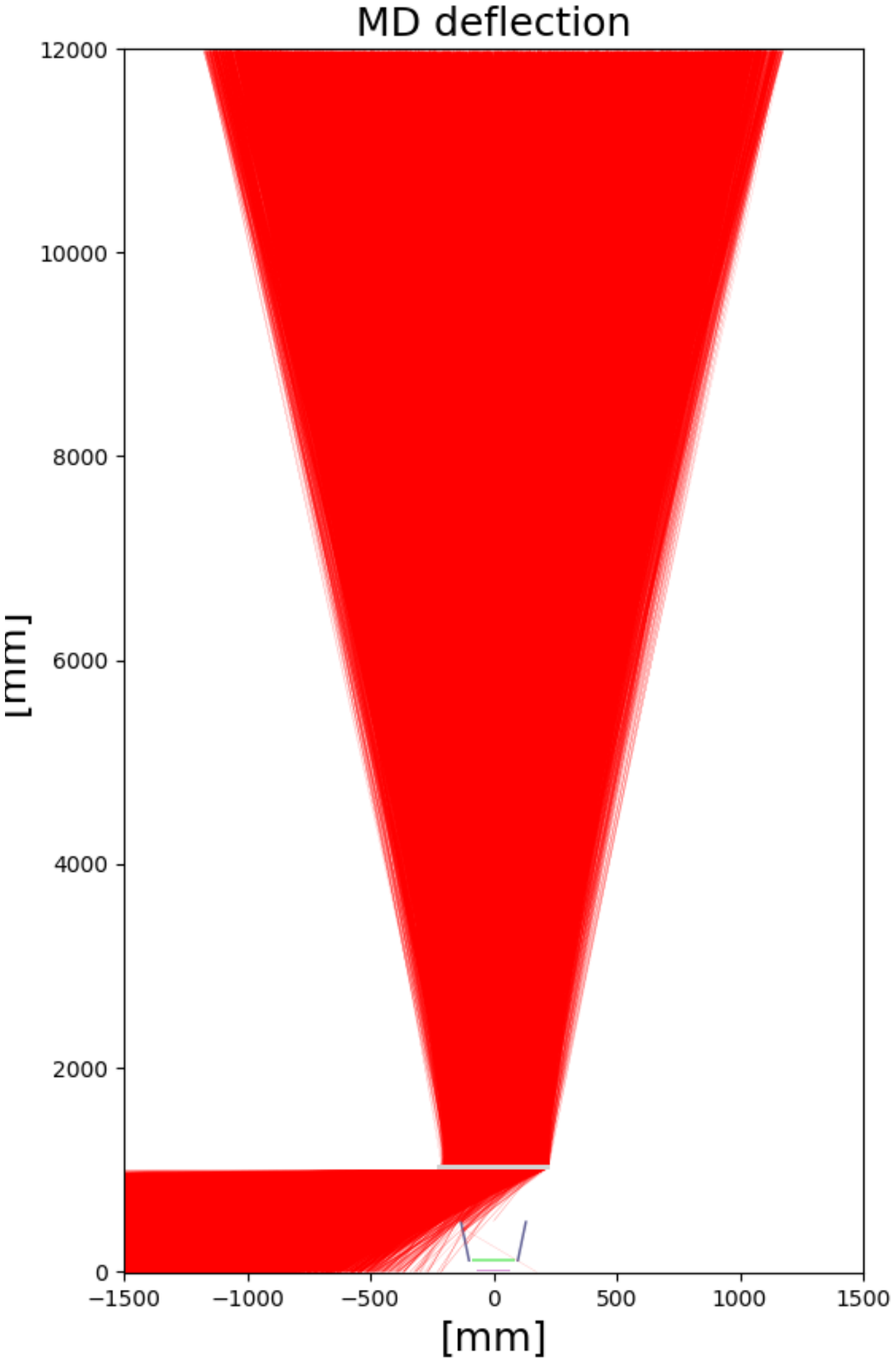}
   \includegraphics[trim={1cm 3cm 3cm 4cm},clip,width=0.49\textwidth]{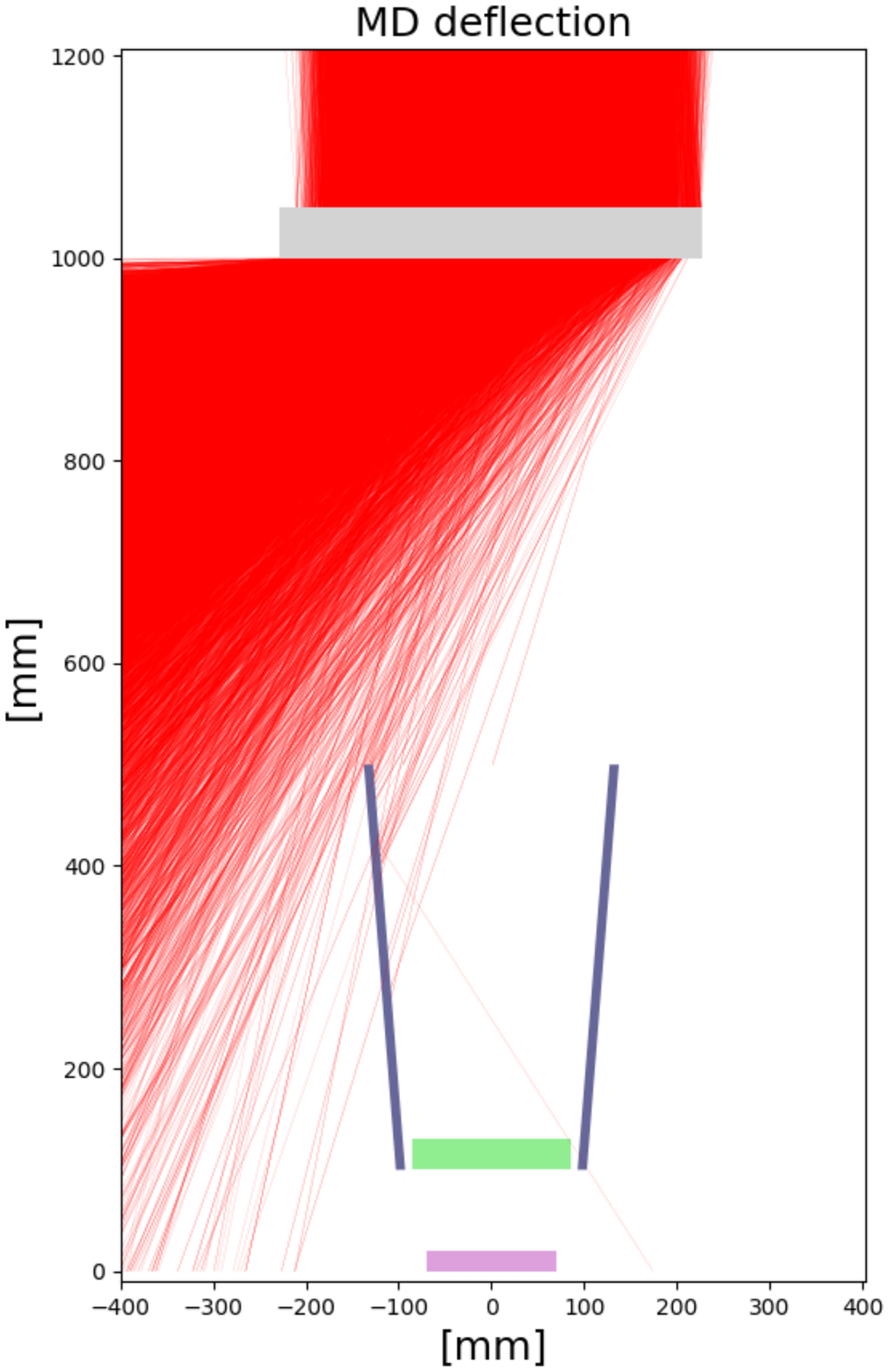}
   \caption 
   { \label{fig:MD}Proton tracks, in red, entering the magnetic diverter and being deflected by its field, for the full ATHENA focal length (left panel) and a zoom on the focal plane region (right panel). The pink, green, dark, and light grey boxes refer to the WFI, the FW optical filters, the baffle, and the diverter internal volume.}
   \end{figure*}
\subsection{Simulating the field}
We assume here a uniform field within the cylinder inner volume: in case of a non-uniform field, the magnetic density simulated here would translate as the minimum value required within the cylinder. A classic Runge-Kutta method for the integration of the particle equation of motion in the field is chosen, with a stepper of 0.1 mm. In Geant4, the curved path of the track is approximated by linear chord segments. The main parameter defining the track segmentation is the distance between the chord and the real curve, set as 1 $\mu m$. The closer is the track to the curved path, the longer is the computation time. The precision in the track segmentation has been increased until no differences were found in the proton deflection. The correct implementation of the diverter field has been tested by simulating a parallel beam of 76 keV protons from the mirror edge towards the opposite diverter side, i.e. the maximum incident angle used in the analytical computation. As expected, protons are deflected just outside the detector assembly (see Fig. \ref{fig:MD_test}, left panel). More specifically, the Geant4 simulation results in protons deflected outside the FW, i.e. the analytical computation is more conservative than the  Monte Carlo simulation. The diverter efficiency highly depends on the required minimum energy to deflect: if 100 keV protons are emitted instead, some of the tracks are able to reach the WFI (Fig. \ref{fig:MD_test}, right panel)
Fig. \ref{fig:MD} shows in red the tracks of protons entering the diverter inner volume, assuming that protons hitting the magnets or the structure surrounding them are fully absorbed. The Geant4 implementation has been tested first by simulating 76 keV protons from the SPO external diameter towards the diverter edge, and checking that they are deflected outside the WFI area.
\\
For the present diverter configuration and simulation statistics, we find that no protons hit the FW or the OBF, and consequently we get a 100\% efficiency in reducing the soft proton induced background on the WFI, for the plasma sheet input regime. A number of background counts below 1, for a 2 -- 7 keV energy range, a 1 second exposure of the current simulation and a WFI sensitive surface of 177.2 cm$^{2}$, would translate into a very much conservative upper limit for the background flux of $1\times10^{-3}$ counts cm$^{-2}$ s$^{-1}$ keV$^{-1}$. Scaling this upper limit for the most probable magnetosheath or solar wind regimes, from 3 to 500 times lower, we can affirm that the proton diverter would ensure a residual background level below the requirement. 
Although some of the protons deflected from the WFI hit the baffle internal walls, no protons or secondary particles are found back-scattering towards the focal plane.


\section{Summary}\label{sec:impl}
The evaluation of soft proton induced X-ray background for future X-ray missions operating outside the radiation belts uses the ATHENA WFI instrument as case study and required the development, for the first time, of an end-to-end simulation of the full interaction chain, from protons entering the optics to the final energy deposits on WFI pixels. The main points of the work presented in this paper can be summarized as follows:
\begin{itemize}
\item soft proton fluxes expected for each plasma regime potentially encountered by a mission orbiting in the L2 or L1 are collected from the AREMBES radiation environment data analysis, for an active magnetosphere and the maximum flux encountered during 90\% of observation time;
\item using Coulomb scattering as physics model for the interaction with the SPO, protons entering the mirror at grazing angle are funneled towards the focal plane;
\item the use of the filter wheel optical filters in addition to the on-chip filters has the effect of lowering by $\sim10$ times the proton flux hitting the WFI and at the same time it increases the average energy of protons inducing background counts, from a dominant range of 10 -- 30 keV to 30 -- 50 keV;
\item the soft proton induced X-ray background on the WFI, in the 2 -- 7 keV energy range, is calculated here to be well above the requirement of $5\times10^{-4}$ counts cm$^{-2}$ s$^{-1}$ keV$^{-1}$ for all the modelled plasma regimes, unless a proton diverter is put in place (see below);
\item a small ($<25\%$) radial variation of the detected surface brightness is observed if only the OBF is present and none if the FW filter is on, in any case well below the radial decrease observed in XMM-Newton soft proton flares;
\item the proton diverter, required to deflect all protons below 76 keV, is simulated as a uniform field of 0.38 T within the Halbach cylinder at 1 m from the WFI;
\item from the present simulation, we have demonstrated that such configuration set-up would ensure a residual focused NXB of the order or well below the requirement, depending on the input proton population.
\end{itemize}

\section{Conclusions and remarks}\label{sec:con}
The efficiency of the proton diverter in minimizing the focused NXB depends on the maximum proton energy to deflect and on the proton incident angle, not on the intensity of the beam. Comparing the impact of different plasma regimes, a harder X-ray spectrum, with more protons at higher energies despite the lower energy integrated flux, could reduce the diverter efficiency. 
Within the framework explored with our simulations we find that the efficiency of the diverter could also decrease if we place additional stopping material in the field of view, which lowers the proton flux at the detector but also increases the maximum proton energy to deflect.
Those lessons are valid for any future mission requiring proton shielding.
\\
From the present work, we identify soft proton scattering by X-ray focusing optics as one of the major sources of unwanted radiation in large effective area X-ray missions as the ATHENA X-ray telescope. The residual background level is proven to be orders of magnitude above the requirements, and a proton diverter is mandatory for the fulfillment of the mission science objectives.
While these results come from a long activity of verification and physics validation of the scattering physics and Geant4 modeling, three main caveats are still present: (i) data from radiation environment observations of low energy protons start from $\sim50-60$ keV, and only extrapolations are possible at lower energies; (ii) laboratory measurements for proton scattering below 200 keV and at low $<1^{\circ}$ scattering angles are still missing, (iii) the proton diverter is assumed to be an ideal Halbach cylinder producing a uniform field within the magnet array.
\\
In terms of accuracy in the input models, requiring a diverter energy threshold above 70 keV would efficiently deflect all protons below, despite their intensity.
Solving once and for all the physics modeling for proton scattering at very long angle, very low energy is one of the aims of the ESA EXACRAD (Experimental Evaluation of Athena Charged Particle Background from Secondary Radiation and Scattering in Optics) project, where experimental measurements are currently ongoing. Geant4 simulations, using both Coulomb scattering as well as alternative new models (e.g. the Remizovich solution), will be compared to data and updates, if needed, will be provided to the community. Geant4 simulations of soft proton scattering by ATHENA mirrors will also be updated if necessary. Finally, in order to achieve a detailed and realistic simulation of the proton diverter, and comparing it with present assumptions, a synergy between the AREMBES and the SIMPOSIuM (SIMulations of Pore Optics in SIlicon and Modelling) ESA projects is currently ongoing, with the aim of using the output of the soft proton scattering simulation as input to a dedicated diverter simulator.

\begin{acknowledgements}
The present work has been supported by the ESA AREMBES project, contract No 4000116655/16/NL/BW. The author VF would like to thank the AAS Scientific Editor for the constant  and kind support in the review process.
\end{acknowledgements}




\bibliographystyle{aasjournal}       
\bibliography{fioretti_bib} 



\end{document}